\documentclass[11pt]{abpaper}
\usepackage{xcolor,tikz,graphicx}
\usepackage{todonotes}
\usepackage{hyperref}
\usepackage{xspace}
\usepackage{relsize}
\usepackage[update,prepend]{epstopdf}
\usepackage{listings}
\usepackage{amsmath,amssymb}
\usepackage{cancel}
\usepackage{url}
\usepackage{booktabs}
\usepackage{bm}
\usepackage{placeins}

\usepackage{nag}

\usepackage{hyperref}
\usepackage{bookmark}

\newcommand{\eV}{\ensuremath{\text{e}\mspace{-0.8mu}\text{V}\xspace}}

\newcommand{\GeV}{\ensuremath{\text{G\eV}}\xspace}
\newcommand{\TeV}{\ensuremath{\text{T\eV}}\xspace}

\usepackage{units}
\newcommand{\half}{\ensuremath{\nicefrac{1}{2}}\xspace}

\newcommand{\pt}{\ensuremath{p_T}\xspace}

\newcommand{\f}[3][]{\ensuremath{f_{#1}(#2,#3)}\xspace}
\newcommand{\xz}{\ensuremath{x\!/\!z}\xspace}

\newcommand{\alphaS}{\texorpdfstring{\ensuremath{\alpha_\mathrm{S}}}{alphaS}\xspace}

\newcommand{\Sherpa}{\textsc{Sherpa}\xspace}
\newcommand{\Alpgen}{\textsc{Alpgen}\xspace}
\newcommand{\MadGraph}{\textsc{MadGraph}\xspace}

\newcommand{\Pythia}{\textsc{Pythia}\xspace}
\newcommand{\Herwig}{\textsc{Herwig}\xspace}

\makeatletter
\DeclareRobustCommand{\kbd}[1]{{\texttt{#1}}}

\g@addto@macro\bfseries{\boldmath}
\makeatother

\usepackage{mathpazo}
\usepackage{microtype}

\newcommand{\img}[2][1.0]{\includegraphics[width=#1\textwidth]{#2}}

\author{\textbf{Andy Buckley}\\[0.1mm] \rmfamily\smaller\emph{School of Physics \& Astronomy, Glasgow University, UK}}
\title{\setstretch{1} Sensitivities to PDFs in parton shower\\ MC generator reweighting and tuning}
\preprints{MCnet-16-02\\ GLA-PPE-2016-01}
\date{\today}

\begin{document}

\begin{abstract}
  Evaluating parton density systematic uncertainties in Monte~Carlo event
  generator predictions has long been achieved by reweighting between the
  original and systematic PDFs for the initial state configurations of the
  individual simulated events. This weighting is now pre-emptively performed in
  many generators, providing convenient weight factors for PDF and scale
  systematics -- including for NLO calculations where counterterms make the
  weight calculation complex. This note attempts a pedagogical discussion and
  empirical study of the consequences of neglecting the effects of PDF
  variations on the beyond-fixed-order components of MC models, and the
  implications for parton shower \& MPI tuning strategies. We confirm that the
  effects are usually small, for well-understood reasons, and discuss the
  connected issue of consistent treatment of the strong coupling between PDFs
  and parton showers, where motivations from physical principles and the need
  for good data-description are not always well-aligned.
\end{abstract}

\section*{Introduction}

Parton shower Monte Carlo (MC) event generators are a key tool for modelling of
collider events beyond fixed perturbative order, and in particular for producing
simulated fully exclusive events which closely resemble those found in
real-world collider experiments. The leading shower MCs include non-perturbative
effects and complex phase-space effects not available to, for example, analytic
resummation calculations.

The price of such realism is the addition of free parameters which must be tuned
to data, an expensive process in both computational and manpower terms. A
cottage industry between shower MC developers and collider experiments has grown
up around this problem~\cite{Buckley:2009bj}, and several generations of tunes,
particularly for the \Pythia shower MC family, have evolved from the LEP and
Tevatron eras up to the present point early into the second run of the
Large Hadron Collider (LHC)~\cite{Skands:2014pea,ATL-PHYS-PUB-2014-021}.

The last ten years of developments in shower MC technology have led to fully
exclusive event generation in which the parton shower (PS), and usually also
non-perturbative modelling aspects such as hadronization and multiple partonic
interactions (MPI), are smoothly matched to matrix element (ME) calculations
significantly improved over the leading-order Born level. These include both
``multi-leg LO'', exemplified by the Alpgen~\cite{Mangano:2002ea},
MadGraph~\cite{Alwall:2011uj}, and Sherpa\,1~\cite{Gleisberg:2008ta} codes, the
``single-emission NLO'' codes such as POWHEG~\cite{Frixione:2007vw} and
(a)MC@NLO~\cite{Frixione:2002bd,Alwall:2014hca}, and the latest generation in
which both modes are combined into shower-matched multi-leg NLO: Sherpa\,2 and
MadGraph5-aMC@NLO~\cite{Frederix:2012ps}.

This computational frontier does not come without downsides: the integration and
efficient sampling of many-leg phase space is slow, as is the computation,
integration and subtraction of the matrix element terms required for finite NLO
calculations. The major benefit of NLO calculations in particular is the
expected robustness of total cross-sections and many differential quantities, to
be explicitly confirmed by constructing envelopes for variation of
renormalization \& factorization scales, and parton density functions (PDFs) in
the calculations. The high CPU cost of the state-of-the-art calculations means
that explicit consistent re-running of the shower-matched simulation for all PDF
and scale variations is unfeasible: instead, internal construction of pre-shower
event weights for each systematic variation has become the standard
approach\footnote{Construction of PDF reweighting factors is more complicated at
  NLO, due to the need for subtraction counter-terms with different initiating
  parton flavours and kinematics. The weight computation is hence done inside
  the NLO matching generator rather than by \textit{post hoc} construction of
  single PDF ratios as for LO events.}.

While computationally necessary, this approach neglects the effect of these
systematic variations on the tuned components of the simulation; in particular
the parton shower (which na\"ively should be configured consistently with the
PDF and scale choices of the matrix element) and the MPI model (which is not
connected to the hard scattering by perturbative QCD, but does display
significant PDF dependence).

The same issues apply to shower MC tuning, for LO
as well as NLO simulations: strictly a different tune should be constructed and
used for each matrix element PDF and scale variation, but again this is
computationally impractical and a single shower MC tune tends to be used for all
matrix element variations.

Evaluating the rationale and empirical support for
this factorized approach, with particular respect to PDF variations, is the
topic of this brief note. This is neither novel ground nor particularly
surprising, but since a review of available literature finds surprisingly little
material on the topic it is hoped that this survey makes a useful contribution.

\section{PDF dependence of parton showers}

We will spend most of this study considering the effect of PDF variations on
initial-state parton shower algorithms, producing initial-state radiation
(ISR). The typical ISR algorithm evolves backwards from the hard process'
incoming legs towards the initial state hadrons, and produces only a few
splittings per event as contrasted with the hundreds of branchings in the
typical final-state radiation (FSR) simulation. These few branchings, however,
are usually high-energy compared to the FSR splittings which produce internal
structure, and they are the dominant mechanism by which the collinear parton
shower formalism can produce additional isolated jets -- albeit at a rate
usually less than in data and in dedicated higher-order matrix element
calculations.

As the matrix element calculation includes PDF factors in its partonic hard
process amplitudes, QCD evolution of the initiating parton legs must correct
those PDF factors for the changes in initial parton flavour and Bj\"orken
momentum fraction produced by the sequential ISR branchings, as well as the
usual strong coupling and splitting function terms of a FSR shower
splitting. The ISR Sudakov form factor (the probability of \emph{no} QCD
splitting in evolution of a parton between scales $q$ and $Q$ is hence
\begin{equation}
  \label{eq:sudakov}
  \Delta^\text{ISR}_b(q^2, Q^2; x) \sim
  \exp\Bigg\{ -\sum_b
  \int_{q^2}^{Q^2} \! \frac{ d\tilde{q}^2 }{ \tilde{q}^2 }
  \int_{q^2_0/\tilde{q}^2}^{1 - q^2_0/\tilde{q}^2} \! dz \, \frac{\alphaS(\tilde{q}^2)}{2\pi}
  \underbrace{ \frac{ x' \! \f[b]{x'}{\tilde{q}^2} }{ x \f[\mspace{-1mu}a]{x}{\tilde{q}^2} } }_{\text{PDF terms}}
  P_{ba}\mspace{-0.5mu}(z, \tilde{q}^2) \Bigg\}
\end{equation}
for 
shower cutoff scale $q_0 \sim 1\;\GeV$, splitting function $P_{ba}$ going (backwards) from
initial parton flavour $a$ to $b$, and the PDF $xf_i$ terms as indicated for $x$
and $x' = \xz$ momentum fractions on the original and new initial splittings
respectively.

The key feature of eq.~\eqref{eq:sudakov} from our perspective is that the PDFs
enter the (non-)splitting probability in a ratio between the same PDF at two
momentum fractions $x$ and $x' = \xz$. Hence a change in hard process PDF which is
unmatched by the same change to the ISR PDF will introduce deviations
proportional to the \emph{double ratio} of the two different PDFs between the
two $x$s. As we shall demonstrate empirically, this double ratio is a more
stable quantity than the bare ratio of two PDFs
and hence the effects are not as large as a pure change of PDFs
in the hard process matrix element. Of course, the PDF effect on a single
Sudakov is not the effect on the whole event, since the splitting is iterated,
but the fact that ISR evolution typically only introduces one or two
initial-state splittings per incoming parton means that this effect is not large
-- and by construction it is limited to the production rates and kinematics of
subleading jets not modelled by the hard process.

The effect of PDF systematics and \alphaS variations on ISR Sudakov factors was
studied a decade ago
~\cite{Gieseke:2004tc}, concluding that the
effects of PDF uncertainties are small compared to other sources of uncertainty,
except in certain isolated high- and low-$x$ regions where PDFs are
unconstrained and their uncertainties inflate. In the sections that follow, we
shall reprise the spirit of that study with new PDF fits, and focus on the PDF
(double) ratios entering the Sudakov terms.

\subsection{PDFs between shower splittings}

\begin{figure}[tp]
  \centering
  \img[0.5]{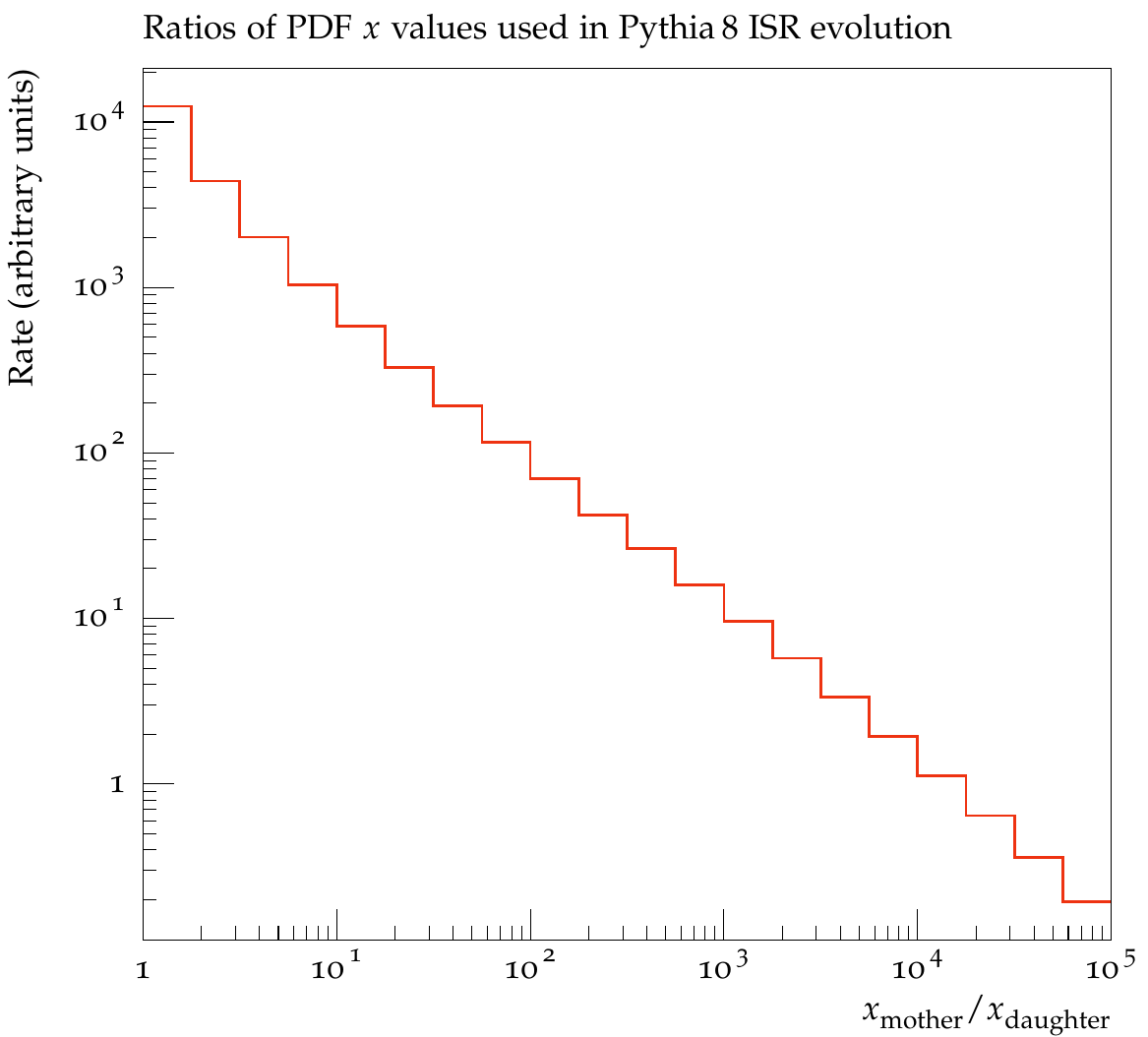}
  \caption{Distribution of $x$ ratios between parent and daughter partons in Pythia\,8 initial-state parton shower evolution.}
  \label{fig:xratio}
\end{figure}

In Figure~\ref{fig:xratio} we show the distribution of $x$ ratios between
initial state shower splittings in Pythia\,8 dijet simulation. It can be seen
that these take the form of a power law, with most splittings close together. To
achieve a balance between the close-together splittings at the low-scale end of
the spectrum and the larger gaps between hard emissions, and to avoid a
proliferation of similar plots, in this section we will use an $x$ ratio of
\half (equivalent to 2 according to the definition in
Figure~\ref{fig:xratio}). Ratio constructions with a factor of 10 produced very
similar results, which are hence elided in the interests of brevity.

\begin{figure}[tp]
  \centering
  \img[0.48]{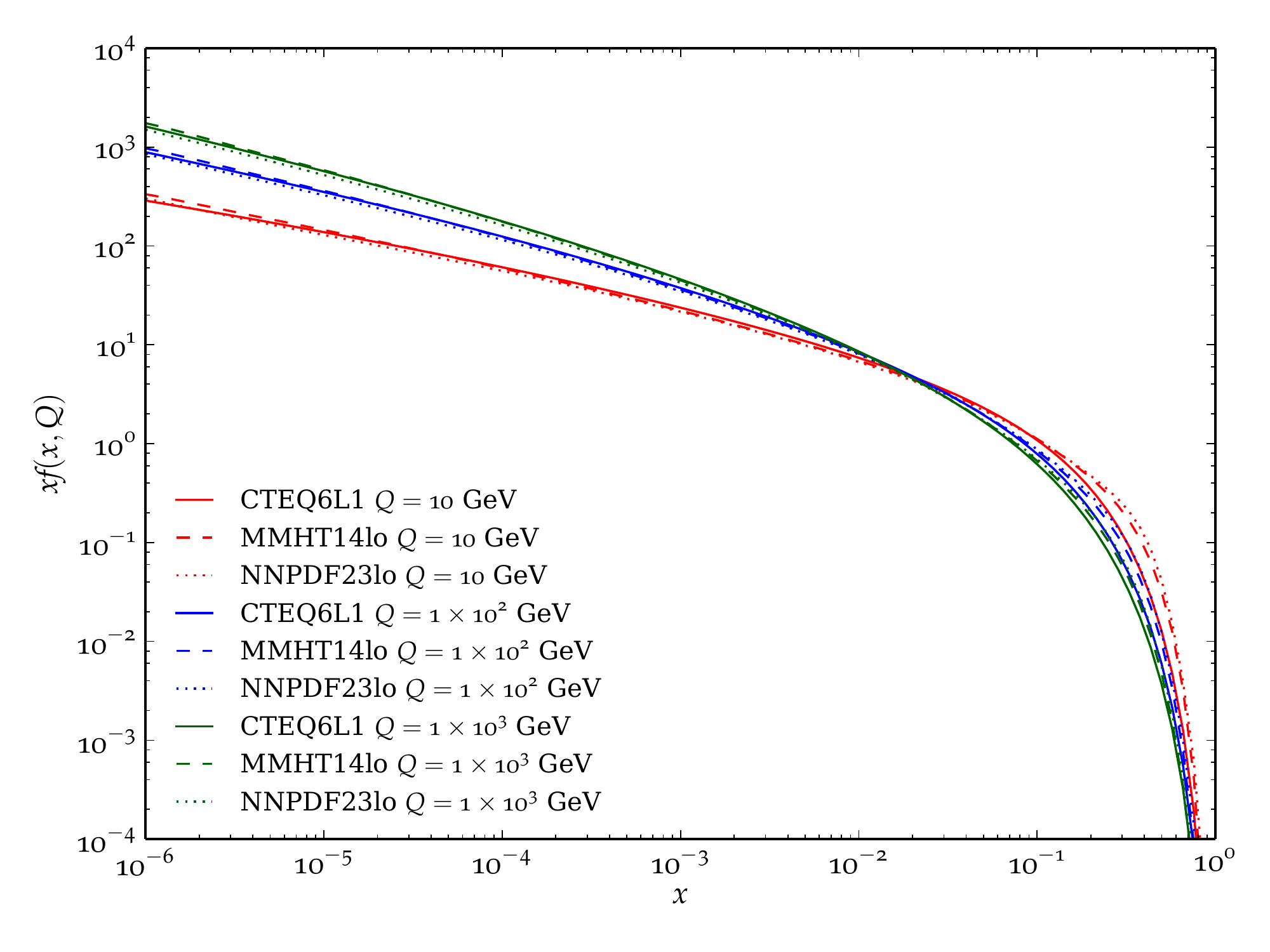}
  \img[0.48]{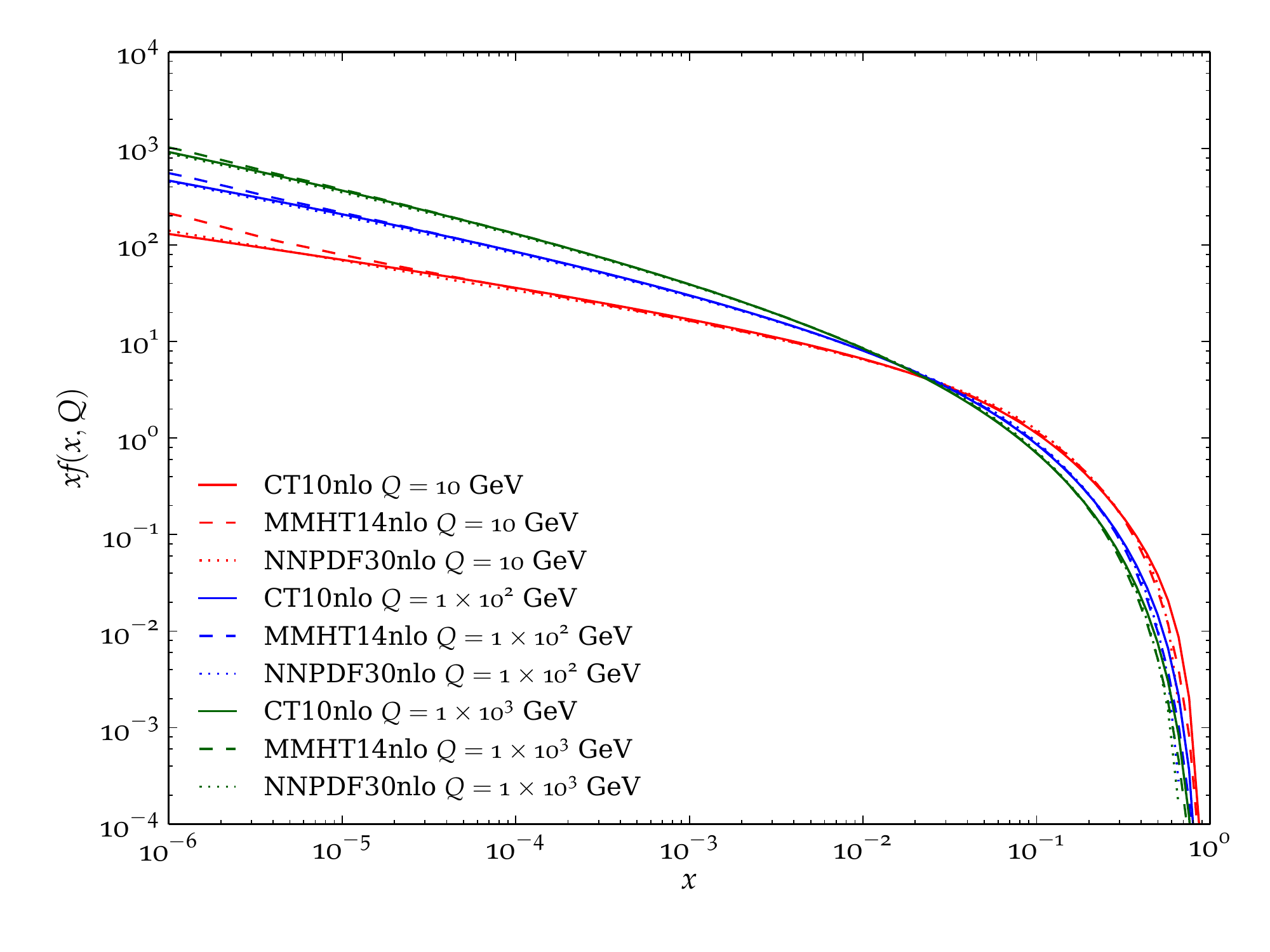}
  \caption{Comparing gluon PDFs from CTEQ, MMHT, and NNPDF at LO (left) and NLO (right).}
  \label{fig:cmppdfs_g}
\end{figure}

\begin{figure}[tp]
  \centering
  \img[0.48]{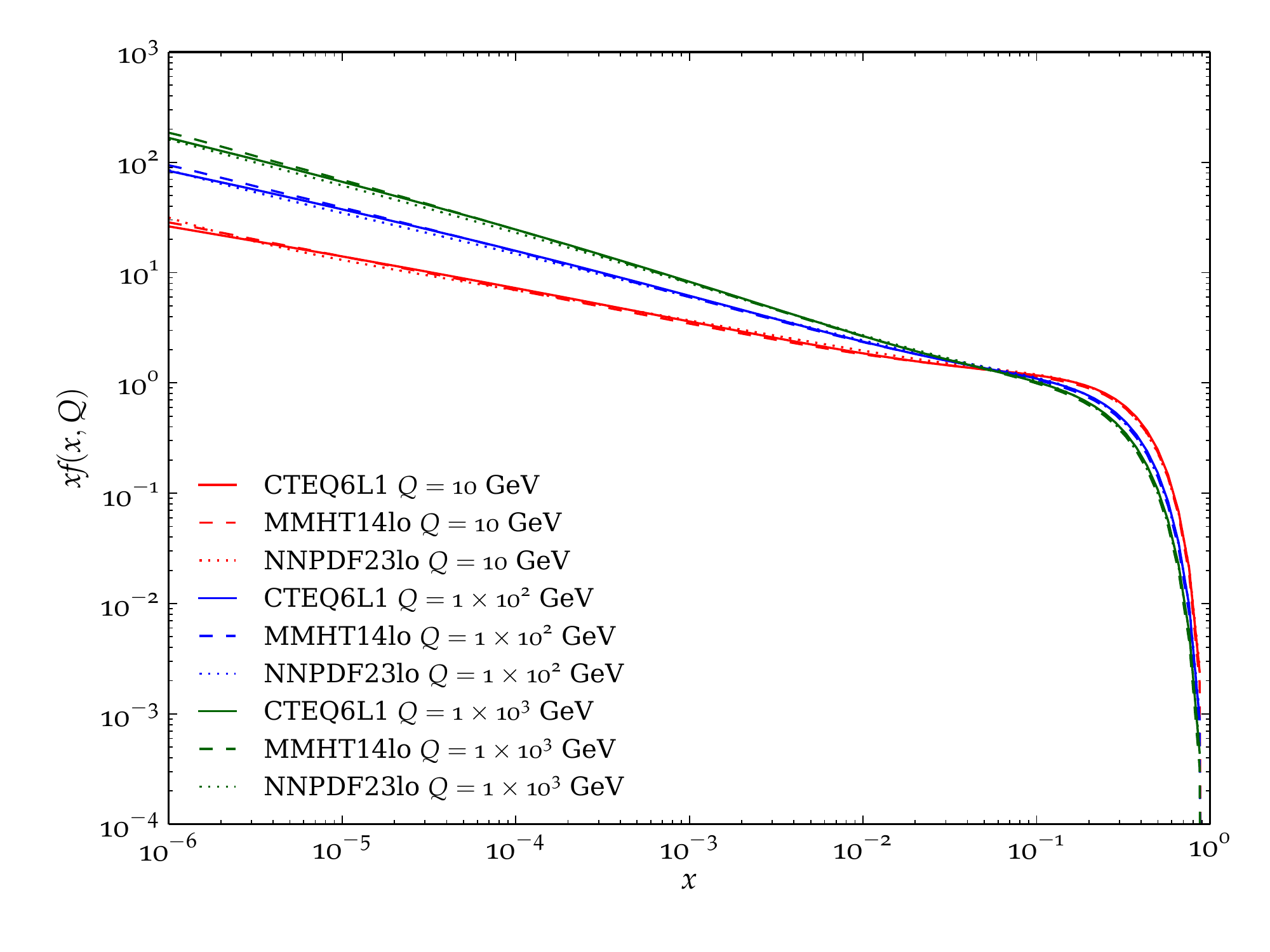}
  \img[0.48]{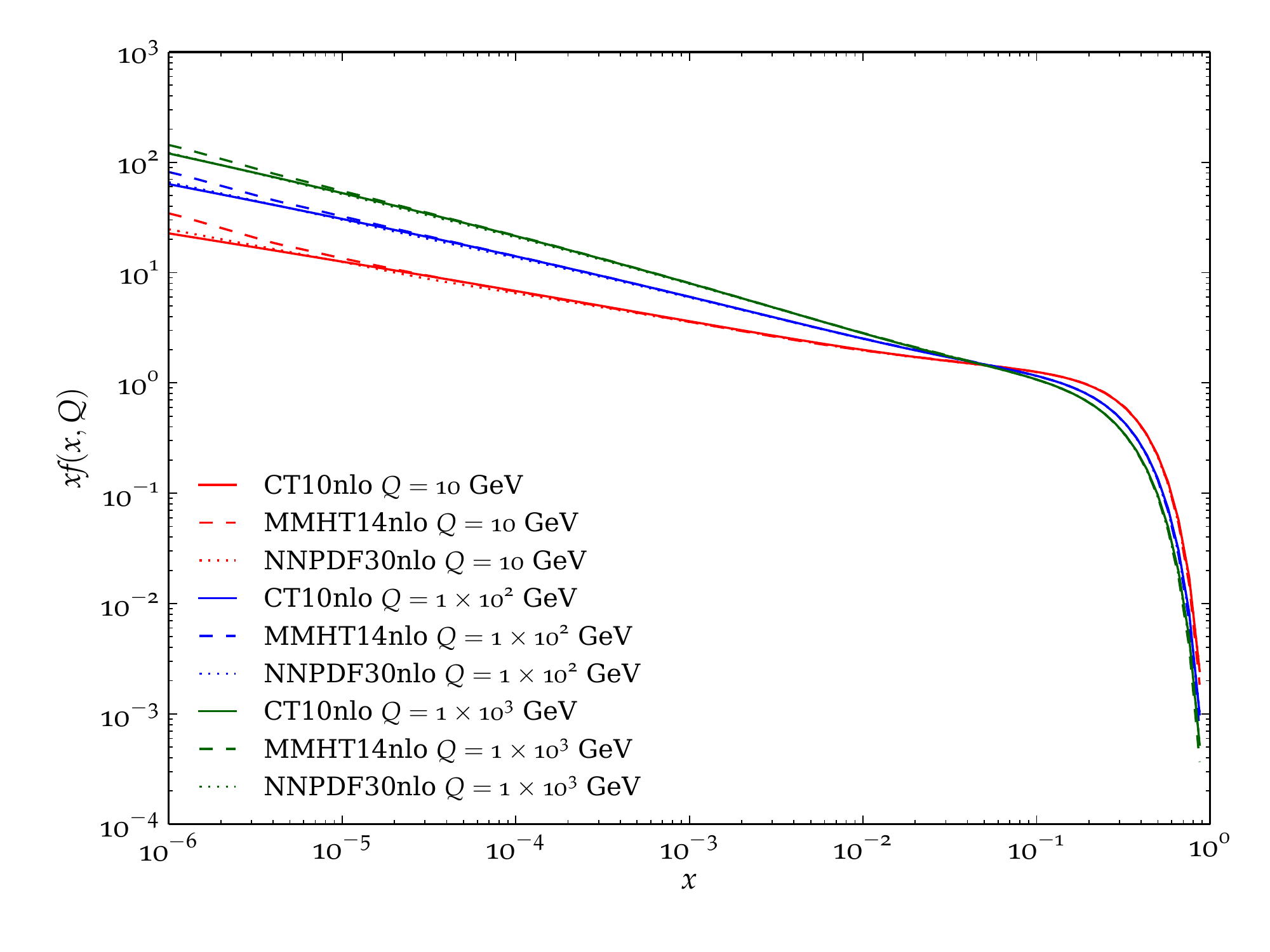}
  \caption{Comparing summed light quark PDFs from CTEQ, MMHT, and NNPDF at LO (left) and NLO (right).}
  \label{fig:cmppdfs_q}
\end{figure}

First, we look at the differences between the latest leading order (LO) and
next-to-leading order (NLO) PDFs from the three major global fit collaborations:
CTEQ~\cite{Guzzi:2011sv}, MMHT~\cite{Harland-Lang:2014zoa} (the latest
incarnation of the MRST / MSTW group), and
NNPDF~\cite{Ball:2013hta,Ball:2014uwa}. These are shown in
Figures~\ref{fig:cmppdfs_g} and~\ref{fig:cmppdfs_q} for gluon PDFs $xf_g$ and
the sum of light (anti)quark PDFs $\sum_{~i\in\text{light}} xf_i$ as functions
of $x$ at several scales from the semi-soft to the very hard\footnote{Adding
  light quark PDFs together for presentation as a single line is implicitly a
  probabilistic interpretation, and hence not strictly valid beyond LO. However,
  it is useful here since negative PDF values only appear in restricted regions
  of phase space, and because a LO interpretation is precisely how they are used
  in parton shower splitting and MPI models.}. These plots were made using the
LHAPDF\,6 parton density library~\cite{Buckley:2014ana}.

We see that the overall scale of deviations between PDF fits are
relatively small -- on these logarithmic scales they are only really visible at
very low values of $x$, where the MMHT PDFs in particular tend to have higher
sea contributions for both quarks and gluons, and for the high-$x$ gluon where
the CTEQ PDFs in particular tend to be at odds with the other two families at
both LO and NLO.

We now look at the ratios of each of these PDFs between different emission
scales, i.e. the PDF ratio term with different $x$ values on numerator and
denominator which appears in eq.~\eqref{eq:sudakov}. We define this ratio for
parton flavour $i$ in PDF $p$ as
\begin{equation}
  \label{eq:ratiodef}
  R^{(p)}_i(x,\tilde{q},z) = \xz \, \f[i]{\xz}{\tilde{q}} \, / \, x \f[i]{x}{\tilde{q}}  .
\end{equation}
%

\begin{figure}[tp]
  \centering
  \img[0.48]{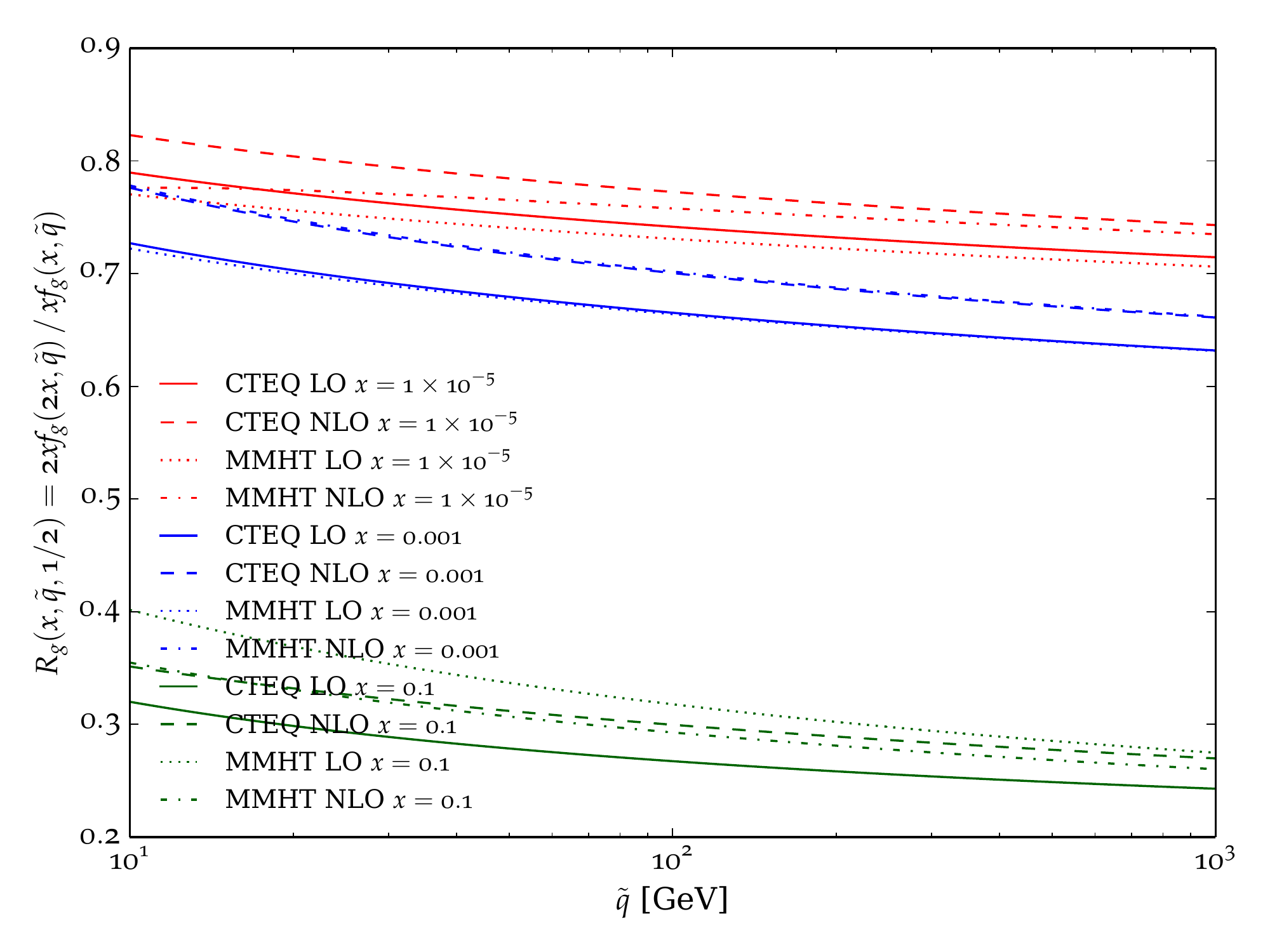}
  \img[0.48]{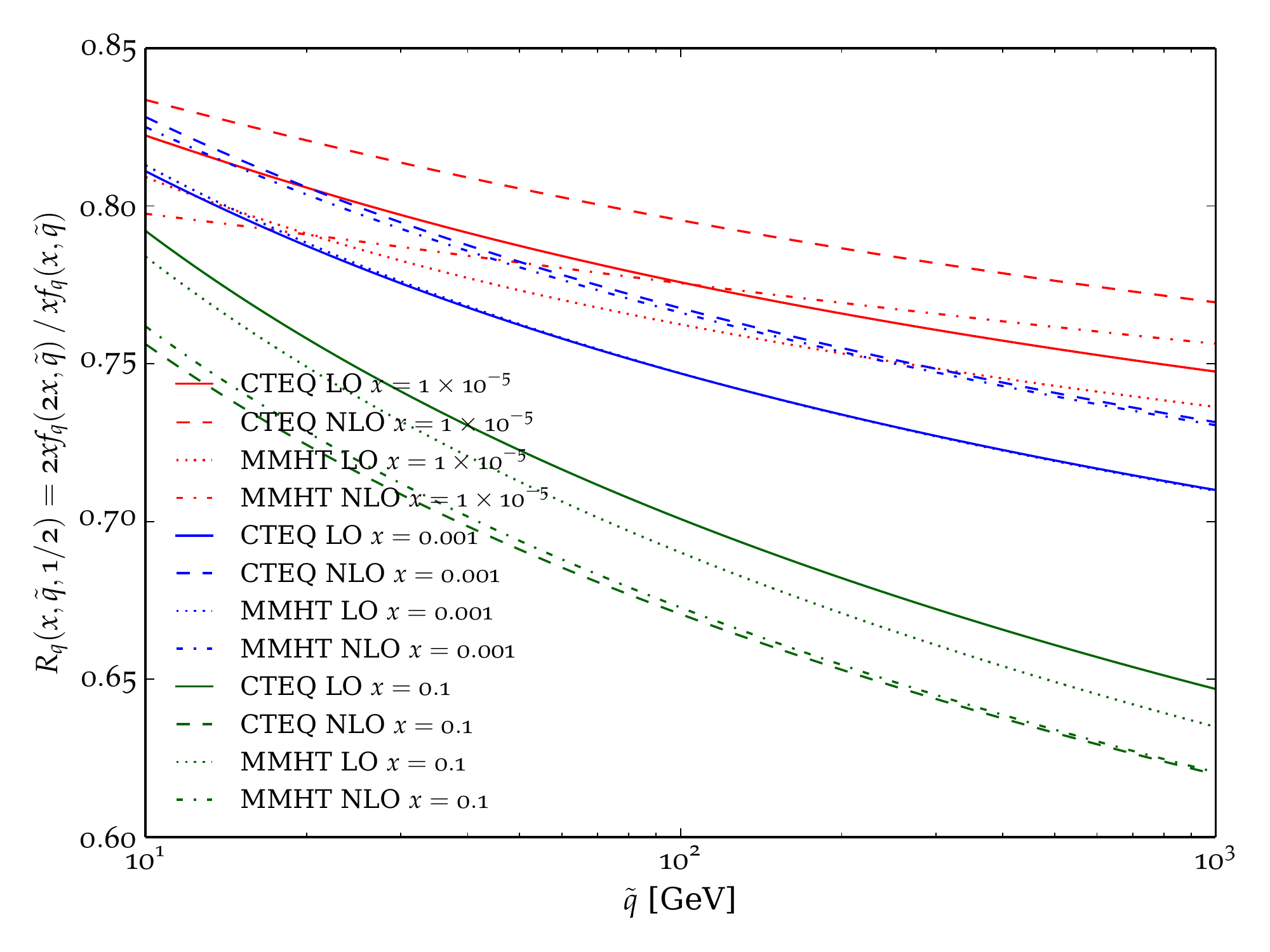}
  \caption{Comparing gluon (left) and light quark (right) PDF ratios between CTEQ/MMHT and LO/NLO, with splitting fraction $z=\half$.}
  \label{fig:sglratios}
\end{figure}

In Figure~\ref{fig:sglratios} we show this ratio as a function of scale
$\tilde{q}$ for the CTEQ and MMHT PDFs which mostly defined the envelope of PDFs
in Figures~\ref{fig:cmppdfs_g} and~\ref{fig:cmppdfs_q}, for a fixed splitting
fraction $z = \half$ -- this nominal $z$ value is taken for illustration, since a
full integration over splitting phase space and splitting functions is best done
in an MC shower generator itself and we shall do just that in
Section~\ref{sec:py8cmps}.

These ratios are clearly quite large, ranging from $\sim 20\%$ deviation from
unity at low values of $x$ and $\tilde{q}$, to 75\% and 35\% deviations at high
$x$ and $\tilde{q}$ for gluons and light quarks respectively. The Sudakov
dependence is stronger in $x$ than in $\tilde{q}$, as expected from
Figures~\ref{fig:cmppdfs_g} and~\ref{fig:cmppdfs_q}: QCD scale evolution of PDFs
is logarithmic, while the $x$ dependence of the PDFs is quite strong, and
becomes more so for the gluon PDF at high-$x$. The potentially large size of
these ratios is not a problem but rather the driver of ISR emission physics; in
the next section we look at the double ratios which reflect the magnitude of
neglected shower effects when reweighting or changing a hard process PDF without
a change of parton shower configuration.

\FloatBarrier

\subsection{PDF double ratios} 

\begin{figure}[tp]
  \centering
  \img[0.48]{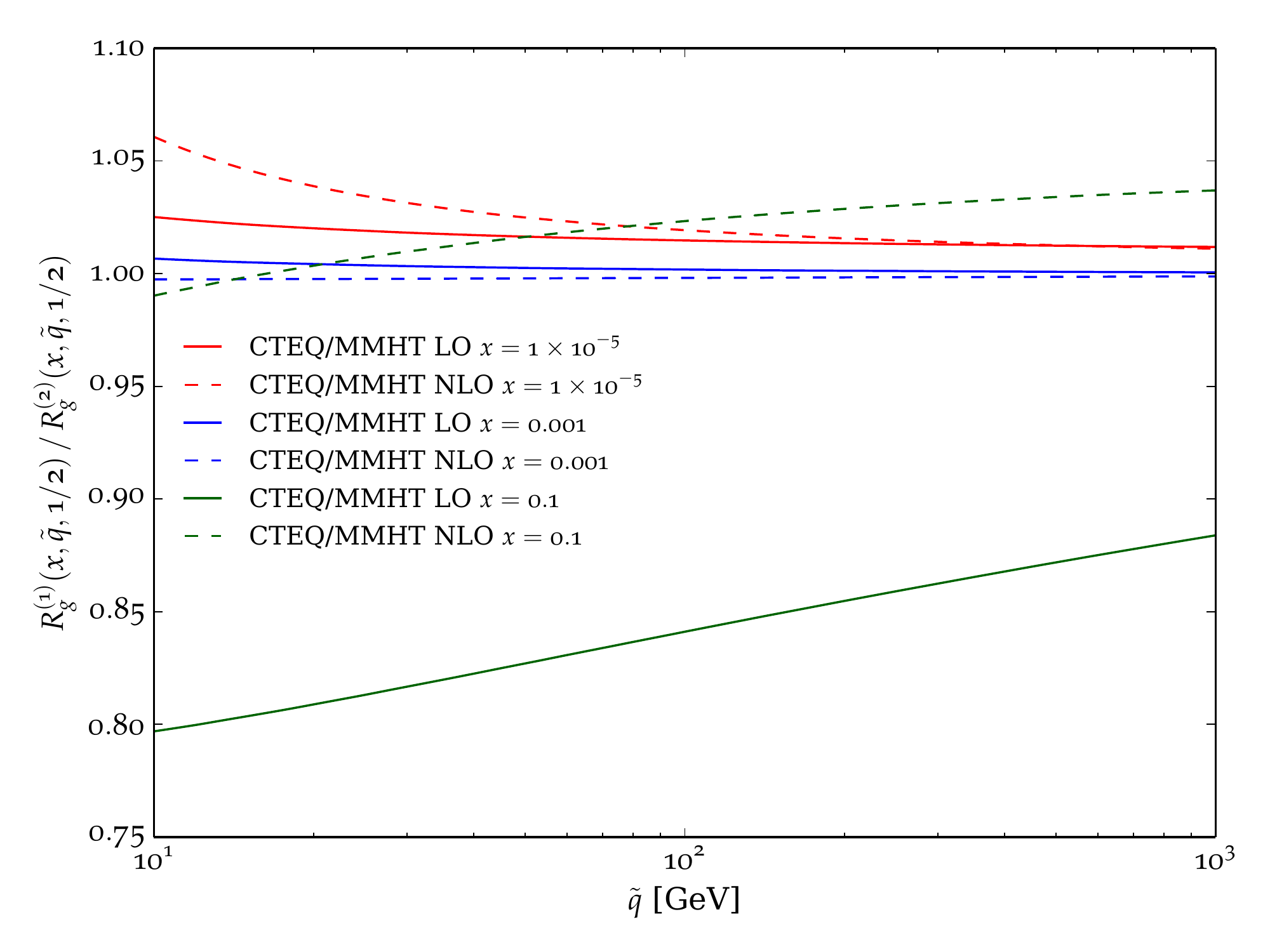}
  \img[0.48]{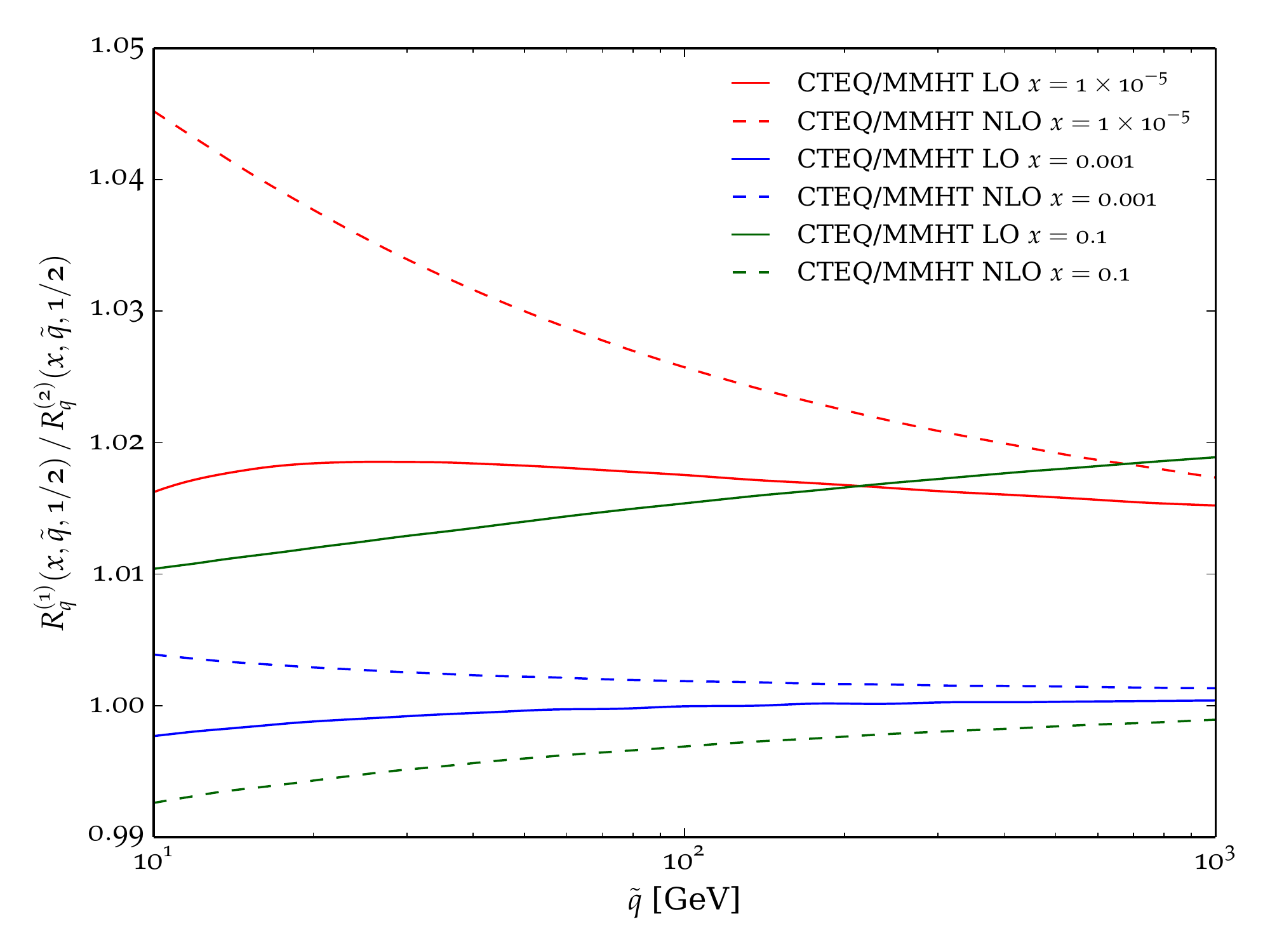}
  \caption{Comparing gluon (left) and light quark (right) PDF double ratios between CTEQ/MMHT, with splitting fraction $z=\half$.}
  \label{fig:dblratio_ct2mm}
\end{figure}

\begin{figure}[tp]
  \centering
  \img[0.48]{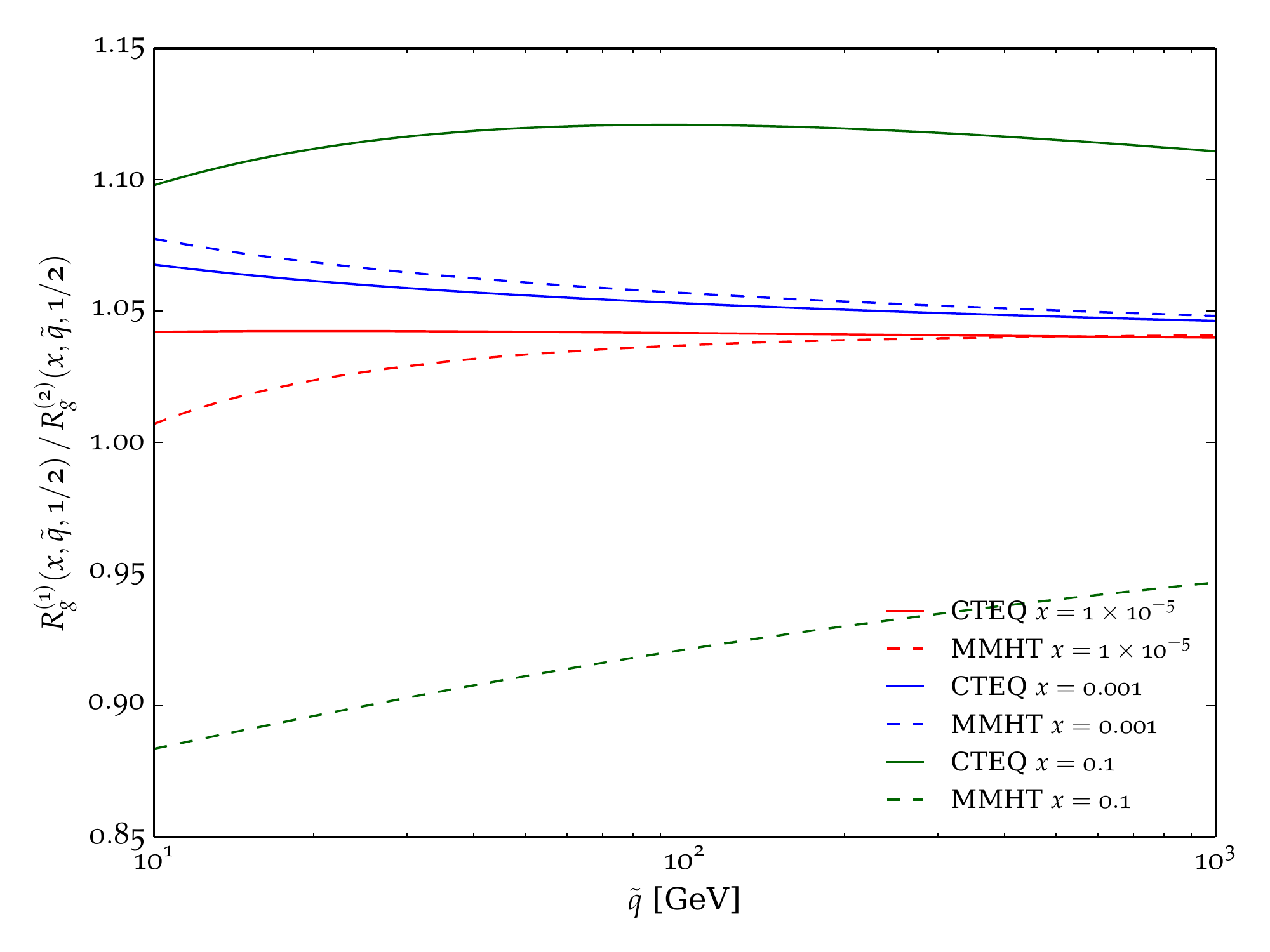}
  \img[0.48]{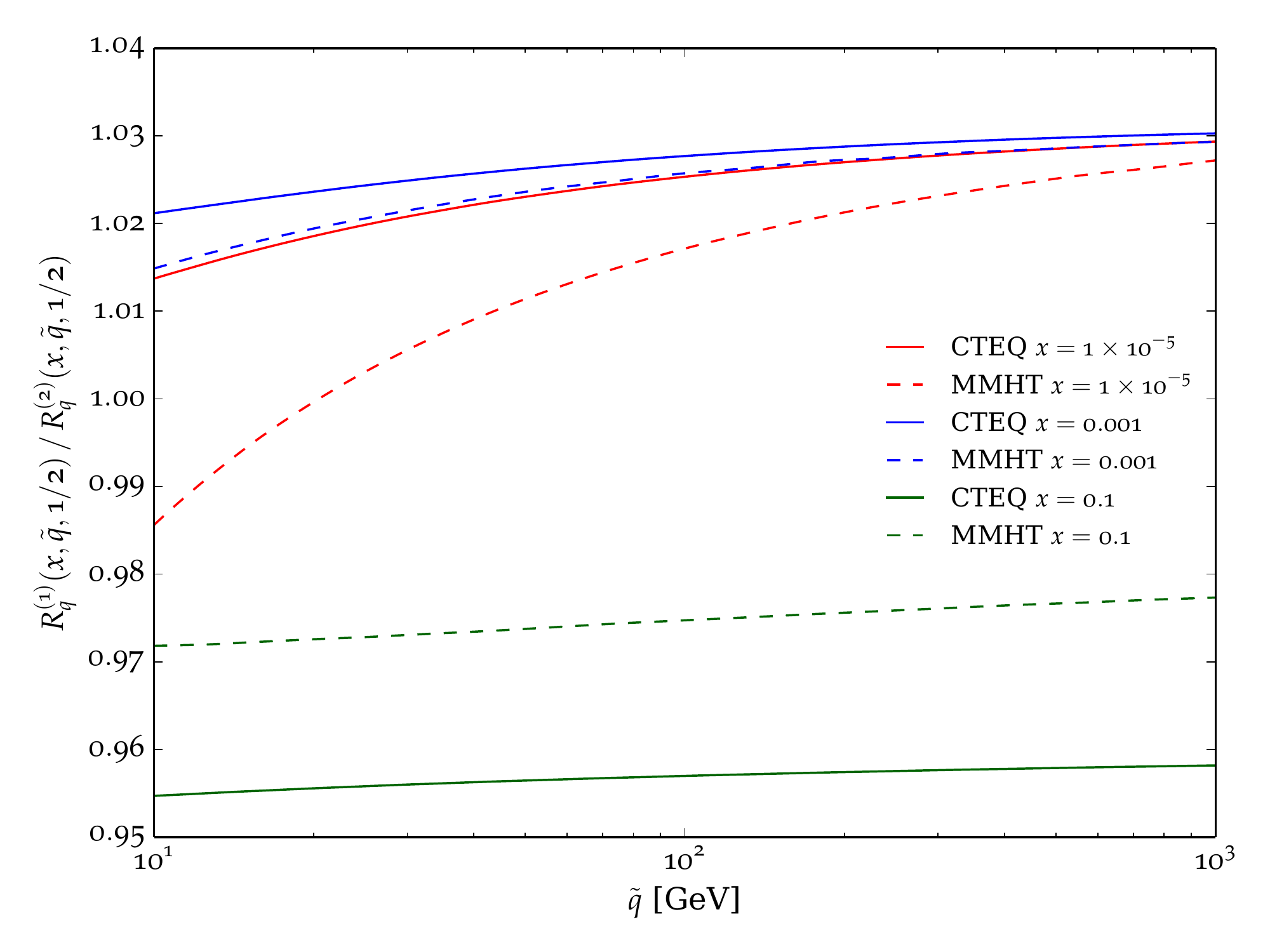}
  \caption{Comparing gluon (left) and light quark (right) PDF double ratios between LO/NLO, with splitting fraction $z=\half$.}
  \label{fig:dblratio_lo2nlo}
\end{figure}

In Figures~\ref{fig:dblratio_ct2mm} and~\ref{fig:dblratio_lo2nlo} we finally see
the ratios of Sudakov form factor PDF terms which correspond to the effect of
switching the parton shower PDF without changes in shower cutoff (or \alphaS,
which can be modified both by use of the \alphaS evolution for that PDF, and by
tuning fudge factors~\cite{Sjostrand:2007gs} and QCD-derived
scalings~\cite{Catani:1990rr}). These are again shown for our \textit{ad hoc}
fixed splitting fraction $z=\half$, and between the CTEQ and MMHT latest LO and
NLO fits.

Figure~\ref{fig:dblratio_ct2mm} shows the double ratios obtained by reweighting
/ switching between the CTEQ and MMHT families. These are typically constrained
to within 5\% of unity, and better than that for most of the range. A mild
exception is seen for reweighting between the NLO PDFs at low $x$ and
$\tilde{q}$, where such PDFs are little constrained; and a very large anomalous
deviation of up to 20\% for high-$x$ LO PDFs, particularly at low scales. It is
expected that these high-$x$ effects will be dealt with in the hard process
rather than the parton shower, and also will use NLO rather than LO PDFs when
matrix elements are available, but the potential effect is worth noting.

Typically shower algorithms, being based on leading order splitting functions,
use leading order PDFs. However, in the case of matching showers to NLO matrix
elements there is a school of opinion that the shower should match the matrix
element in PDF choice to avoid discontinuities across the ME/PS hand-over. An
argument has also been made for using NLO PDFs in leading-order multi-leg matrix
elements on the basis that multi-leg MEs include ISR evolution effects which
would be absorbed into the fitting of an LO PDF, and this has similarly been
used to argue that LO matching would require use of NLO shower PDFs. We may ask
whether it is ``allowed'' to reweight between LO and NLO PDFs, and this is
addressed in Figure~\ref{fig:dblratio_lo2nlo}: larger effects up to $\pm 10\%$
are seen in LO/NLO gluon PDF reweighting of a $z=\half$ Sudakov than in LO or NLO
reweighting between PDF families, but the light quarks are still within a few
percent of unity.

In most typical use-cases the effect of either neglecting shower PDF effects in
reweighting, or of using the same shower tune with different PDFs, is hence
expected to be on the order of 5--10\%, comparable to the typical systematic
uncertainty of shower algorithms (typically evaluated by variation of parton
shower starting scales and \alphaS evolution). This justifies the usual approach
of neglecting explicit shower PDF effects, effectively absorbing them into
shower systematics instead, since the effort required for an explicit evaluation
would be disproportionate to the improvement in predictivity or uncertainty
coverage.

\section{PDF effects on parton showered observables}
\label{sec:py8cmps}

In this section we study the real-world effect of PDF changes in the Pythia\,8
shower MC generator~\cite{Sjostrand:2007gs}, for two important processes: inclusive jet
production, and $W$+jet production in the Run~2 LHC $pp$ configuration with
$\sqrt{s} = 13\,\TeV$.

Pythia\,8 allows separate PDFs to be used for the hard (matrix element) and soft
(parton showers \& MPI) components of the event simulation, so to compare the
effects of PDF changes in the soft modelling we use the default Monash~2013 $pp$
tune~\cite{Skands:2014pea} and fix the hard process PDF to
NNPDF\,2.3\,LO~\cite{Ball:2013hta}, then change the soft PDF to the
CTEQ6L1~\cite{Pumplin:2002vw} and MMHT2014\,LO~\cite{Harland-Lang:2014zoa}
leading order fits, as well as the NNPDF\,3.0\,NLO~\cite{Ball:2014uwa},
CT10\,NLO~\cite{Guzzi:2011sv} and MMHT2014\,NLO~\cite{Harland-Lang:2014zoa}
next-to-leading-order central PDFs. While use of NLO PDFs in soft simulation is
discouraged by MC experts~\cite{Sjostrand:2014mcpdf}, it is not unknown and we
believe that an empirical demonstration of the effects is useful.

\begin{figure}[tp]
  \centering
  \img[0.48]{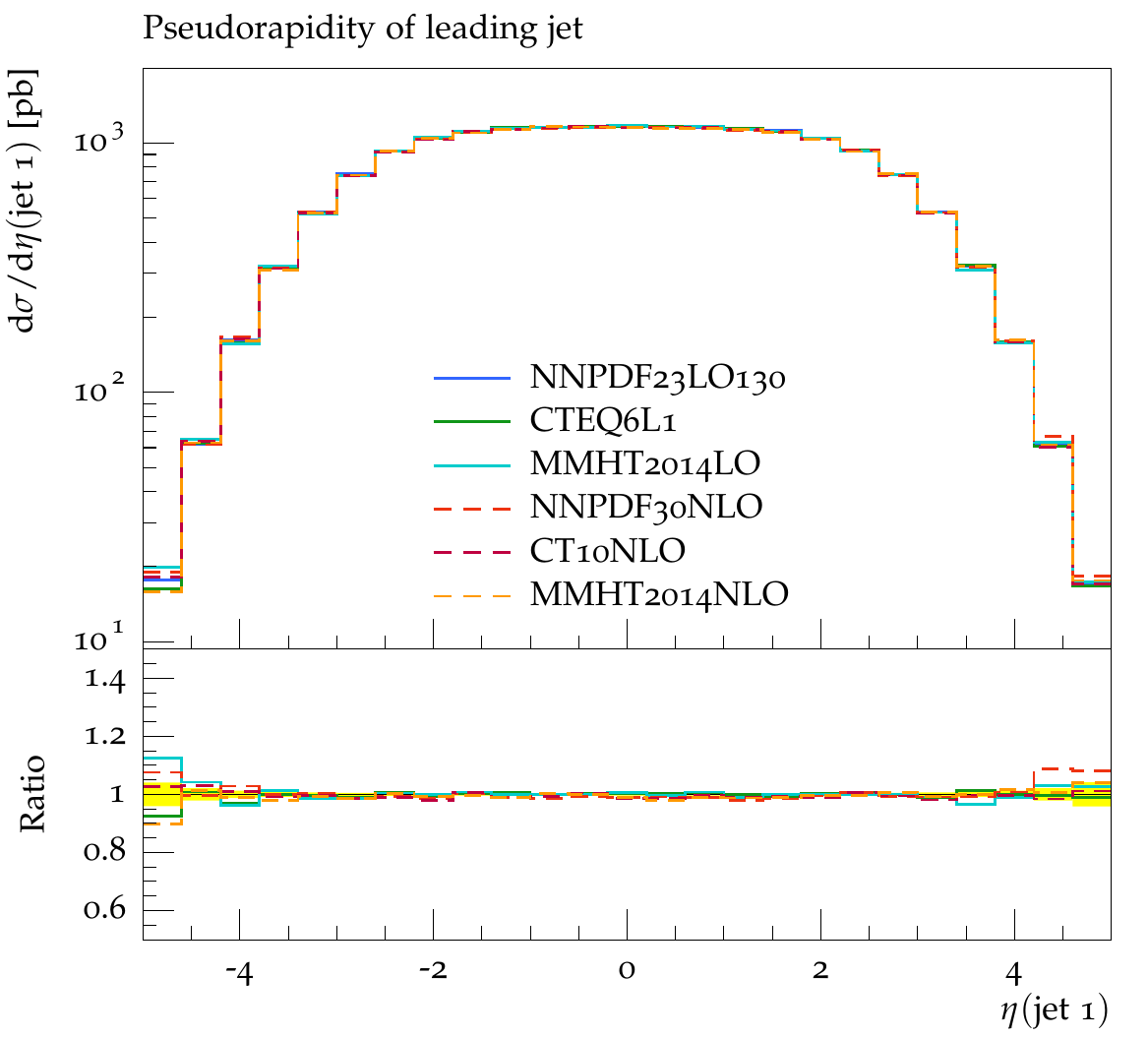}
  \img[0.48]{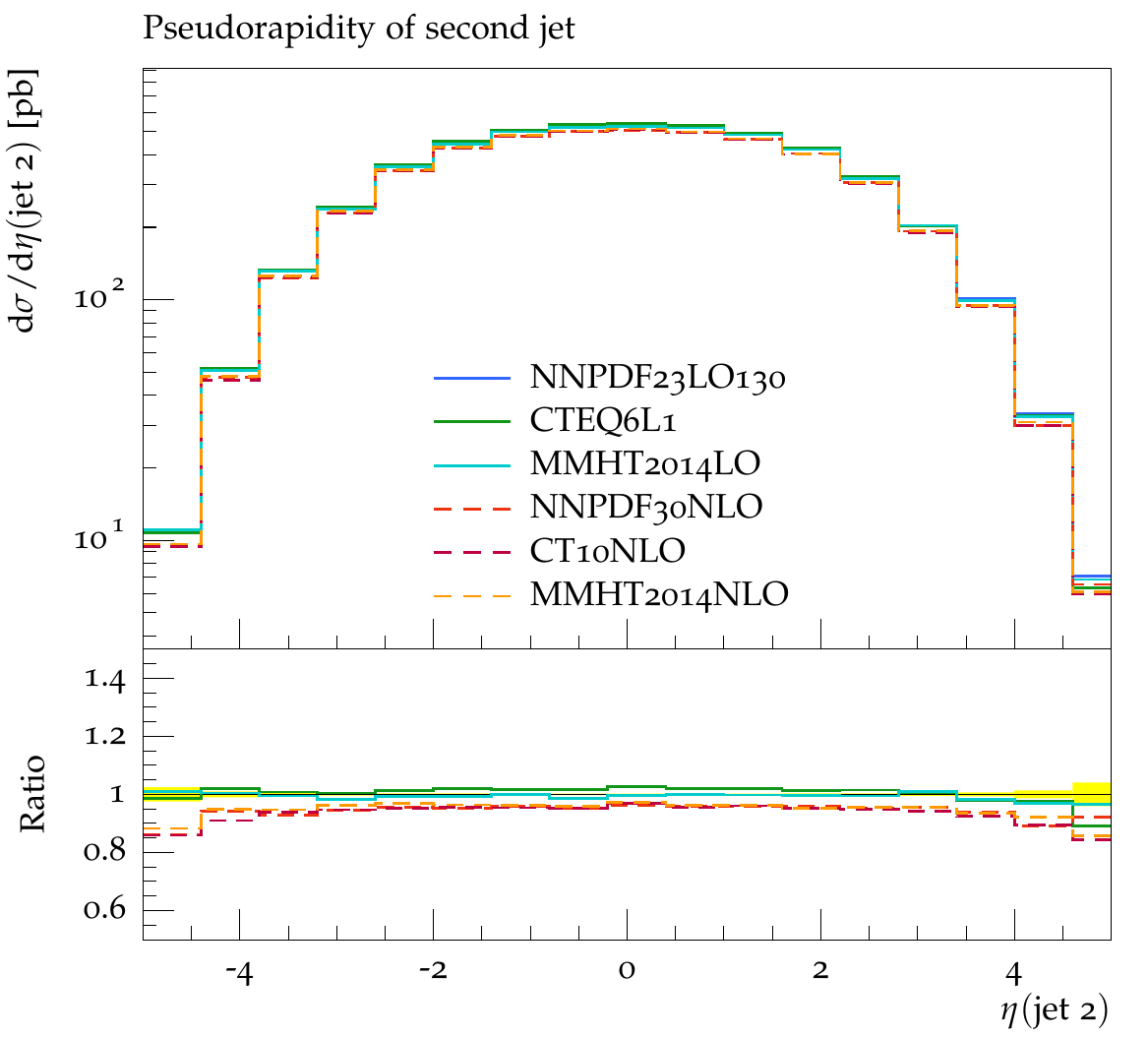}\\[1em]
  \img[0.48]{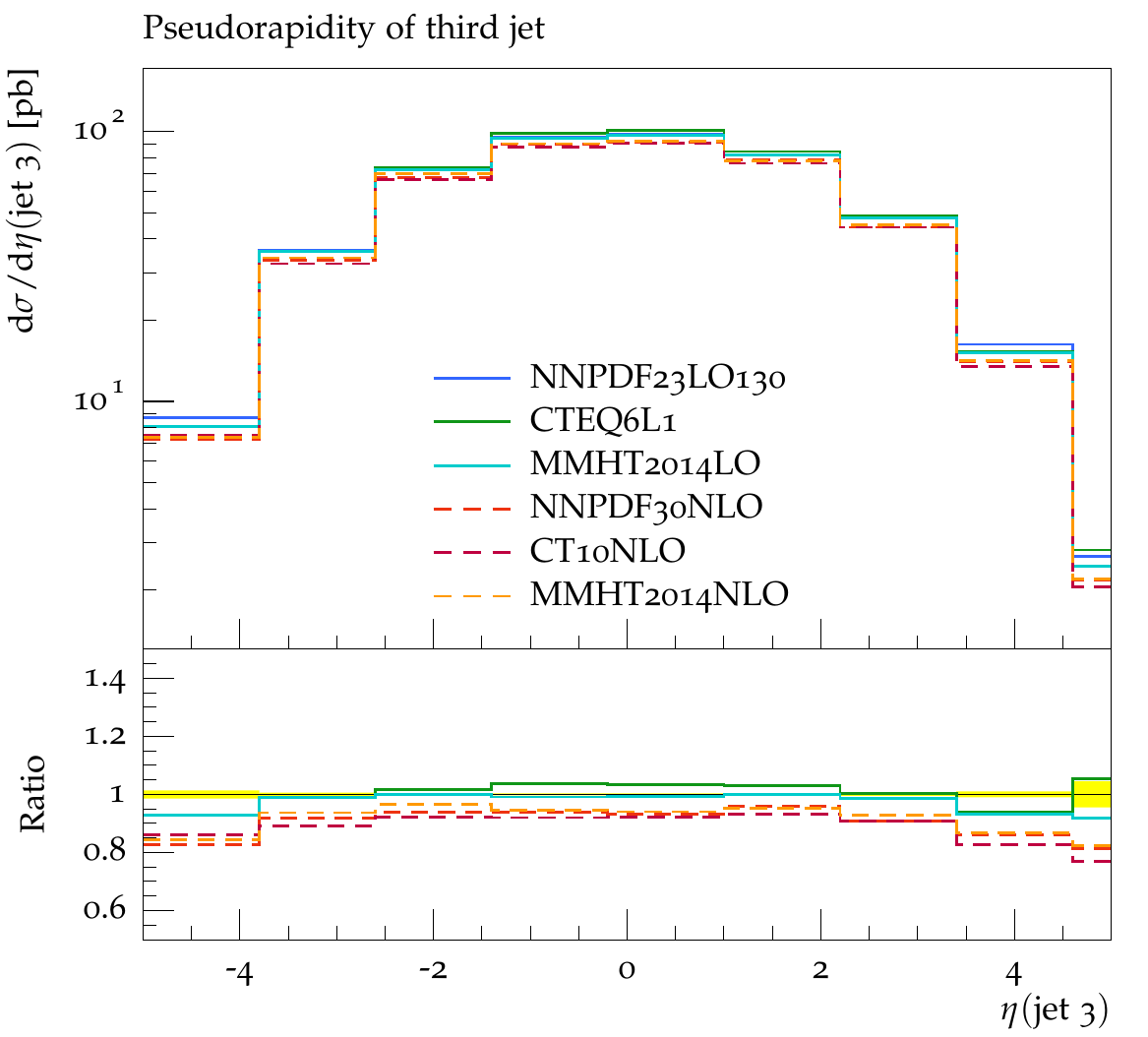}
  \img[0.48]{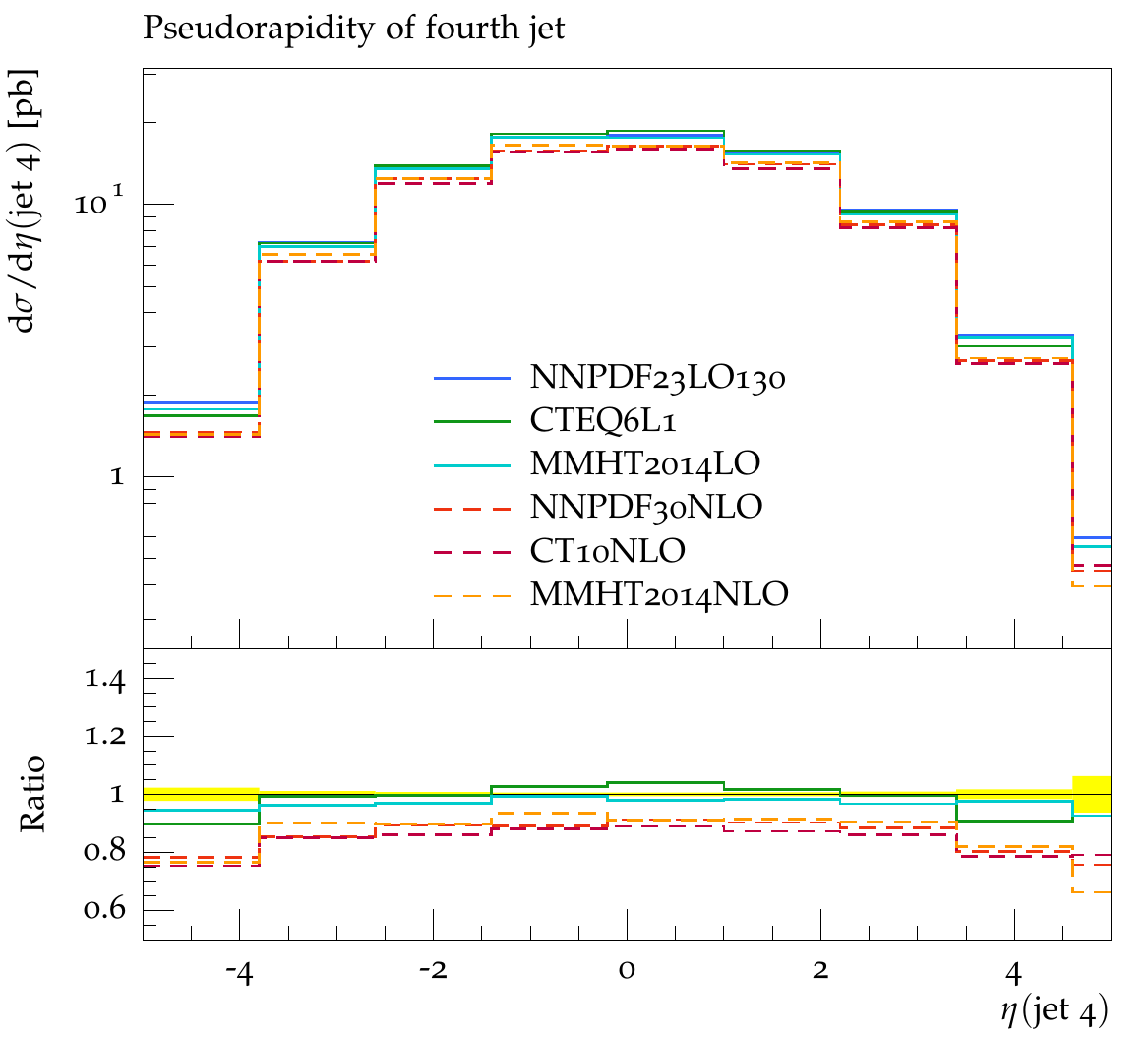}
  \caption{\Pythia8 jet $\eta$ distributions.}
  \label{fig:py8jets2eta}
\end{figure}

\begin{figure}[tp]
  \centering
  \img[0.48]{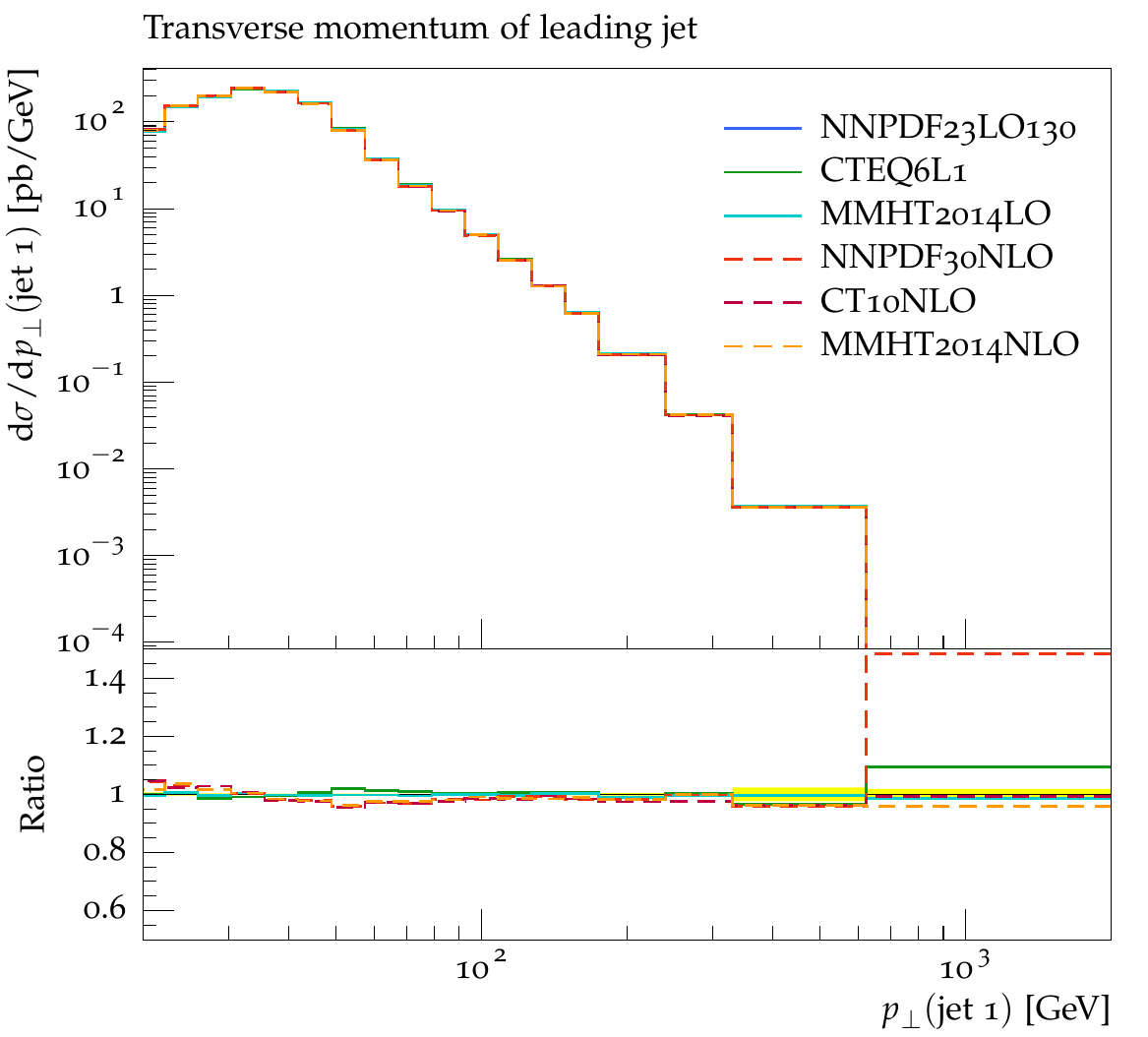}
  \img[0.48]{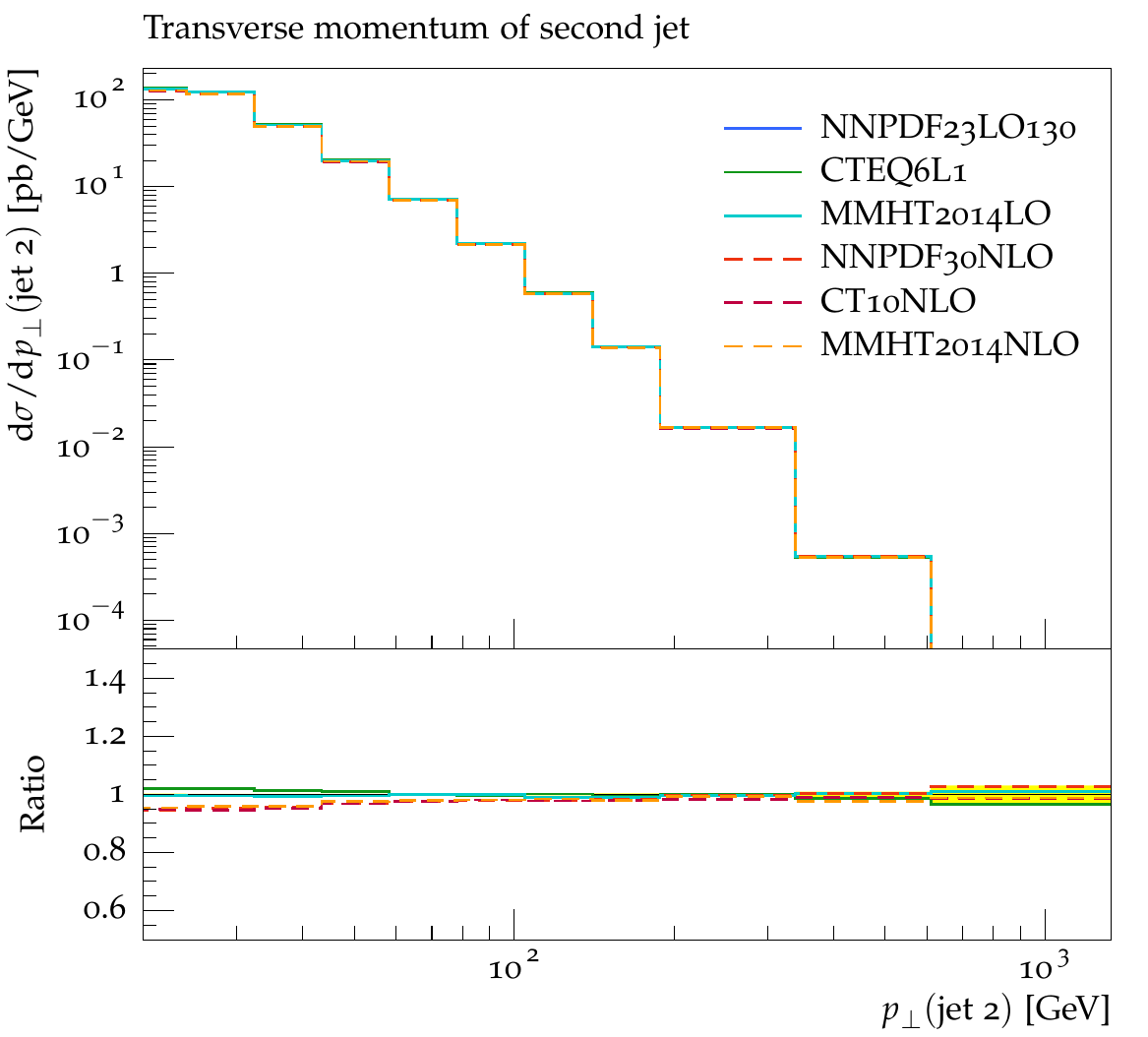}\\[1em]
  \img[0.48]{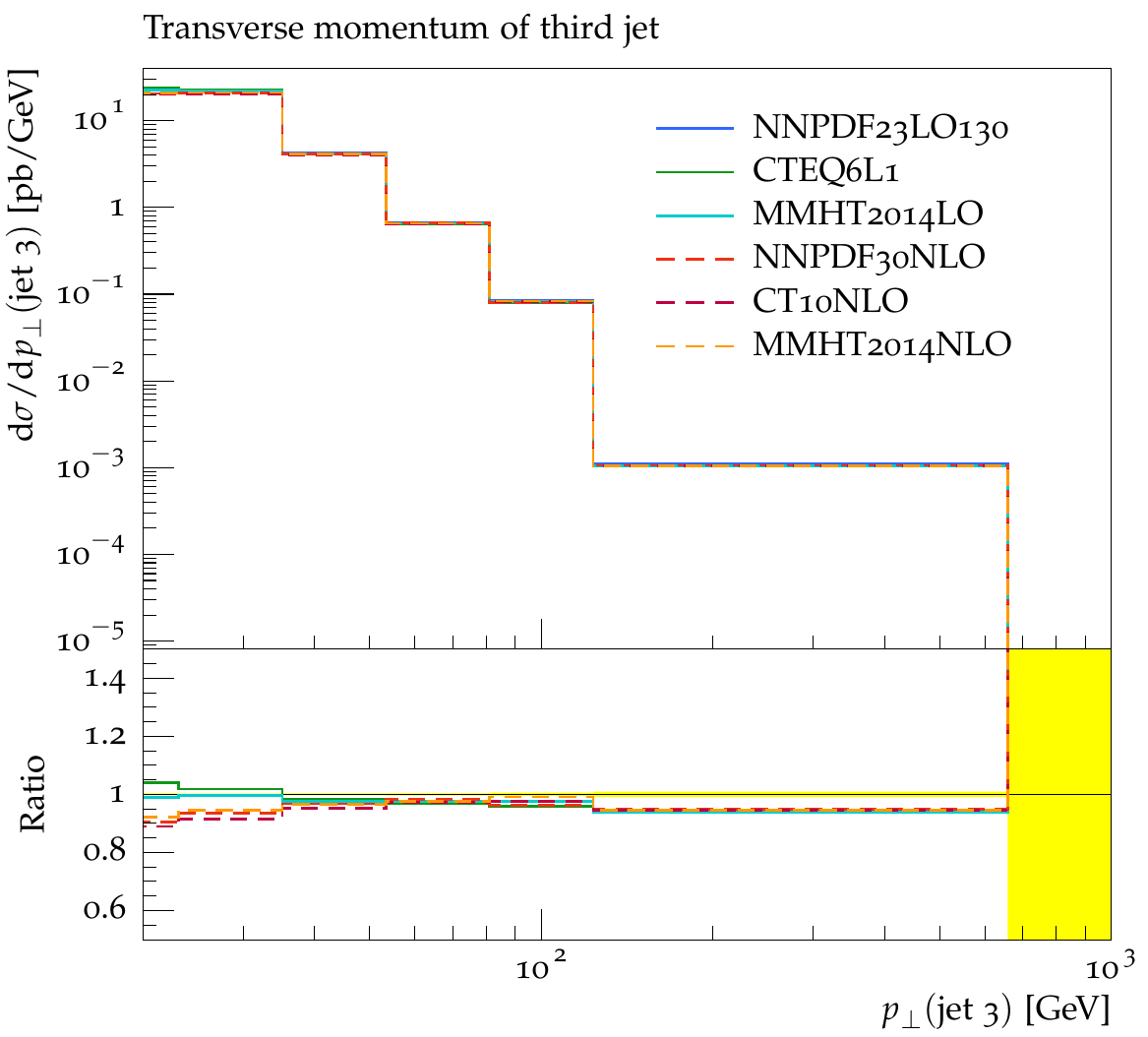}
  \img[0.48]{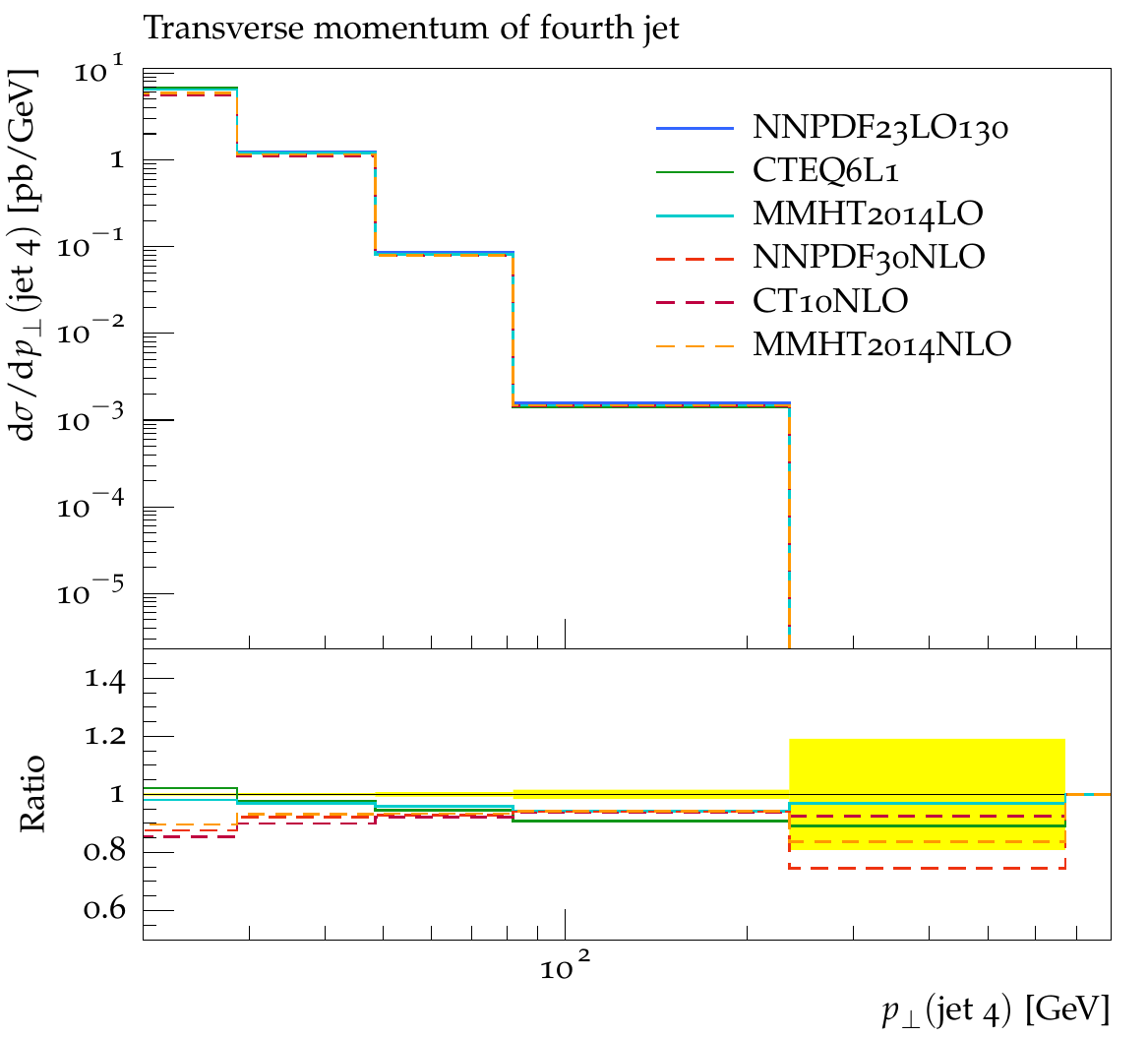}
  \caption{\Pythia8 jet $\pt$ distributions.}
  \label{fig:py8jets2pt}
\end{figure}

\begin{figure}[tp]
  \centering
  \img[0.48]{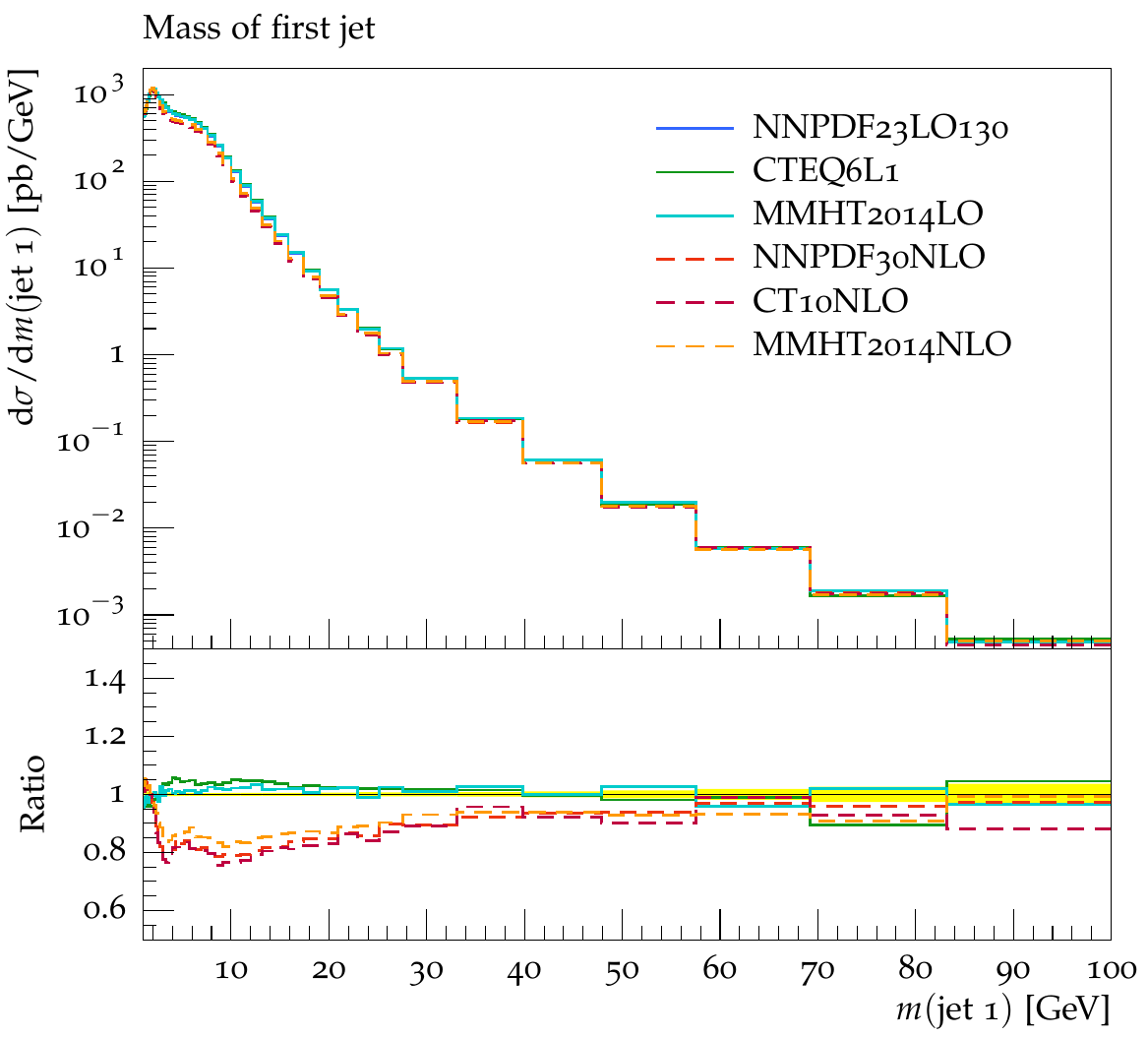}
  \img[0.48]{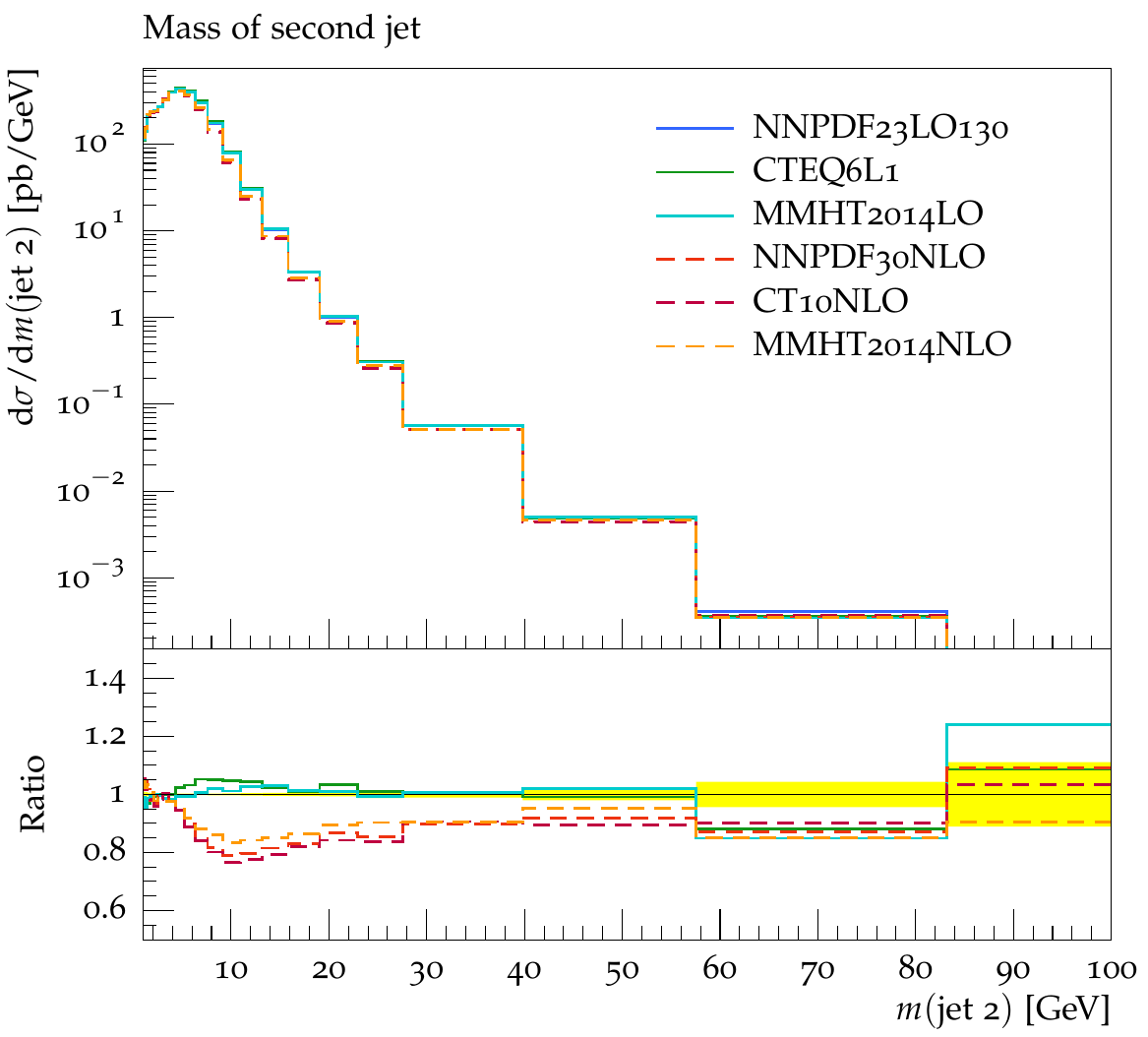}\\[1em]
  \img[0.48]{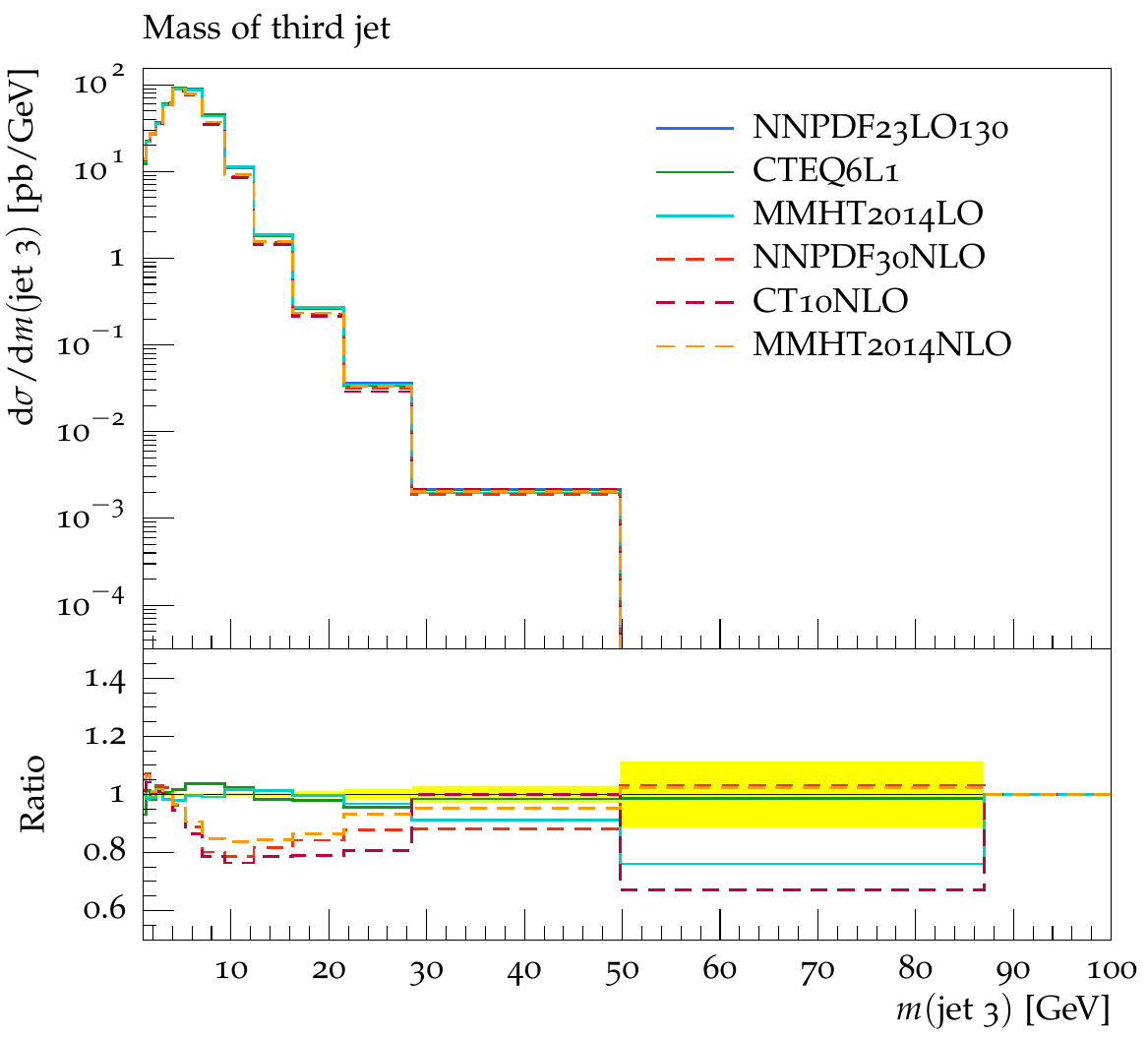}
  \img[0.48]{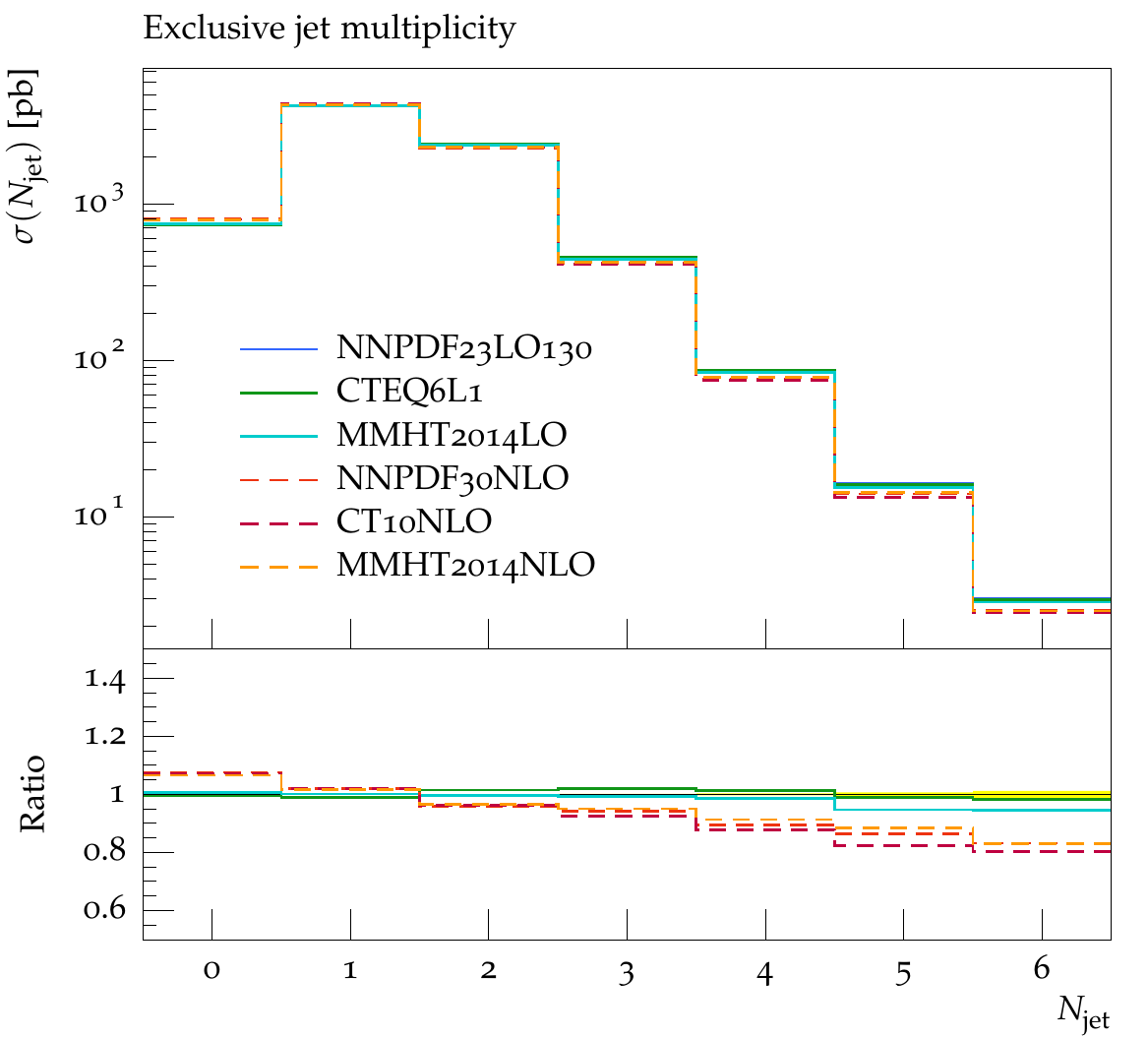}
  \caption{\Pythia8 jet mass and $N_\mathrm{jet}$ distributions.}
  \label{fig:py8jets2mass}
\end{figure}

\begin{figure}[tp]
  \centering
  \img[0.48]{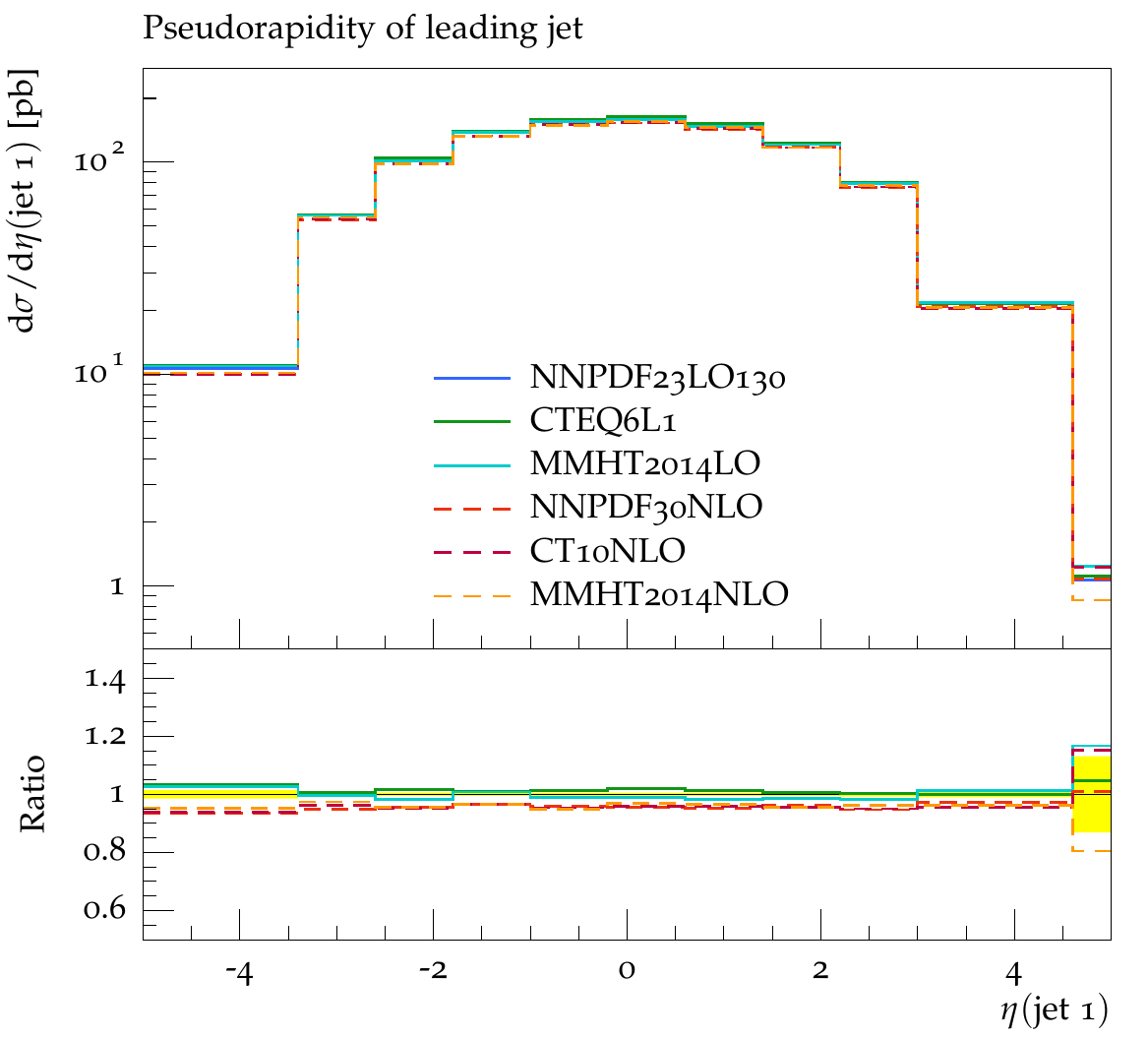}
  \img[0.48]{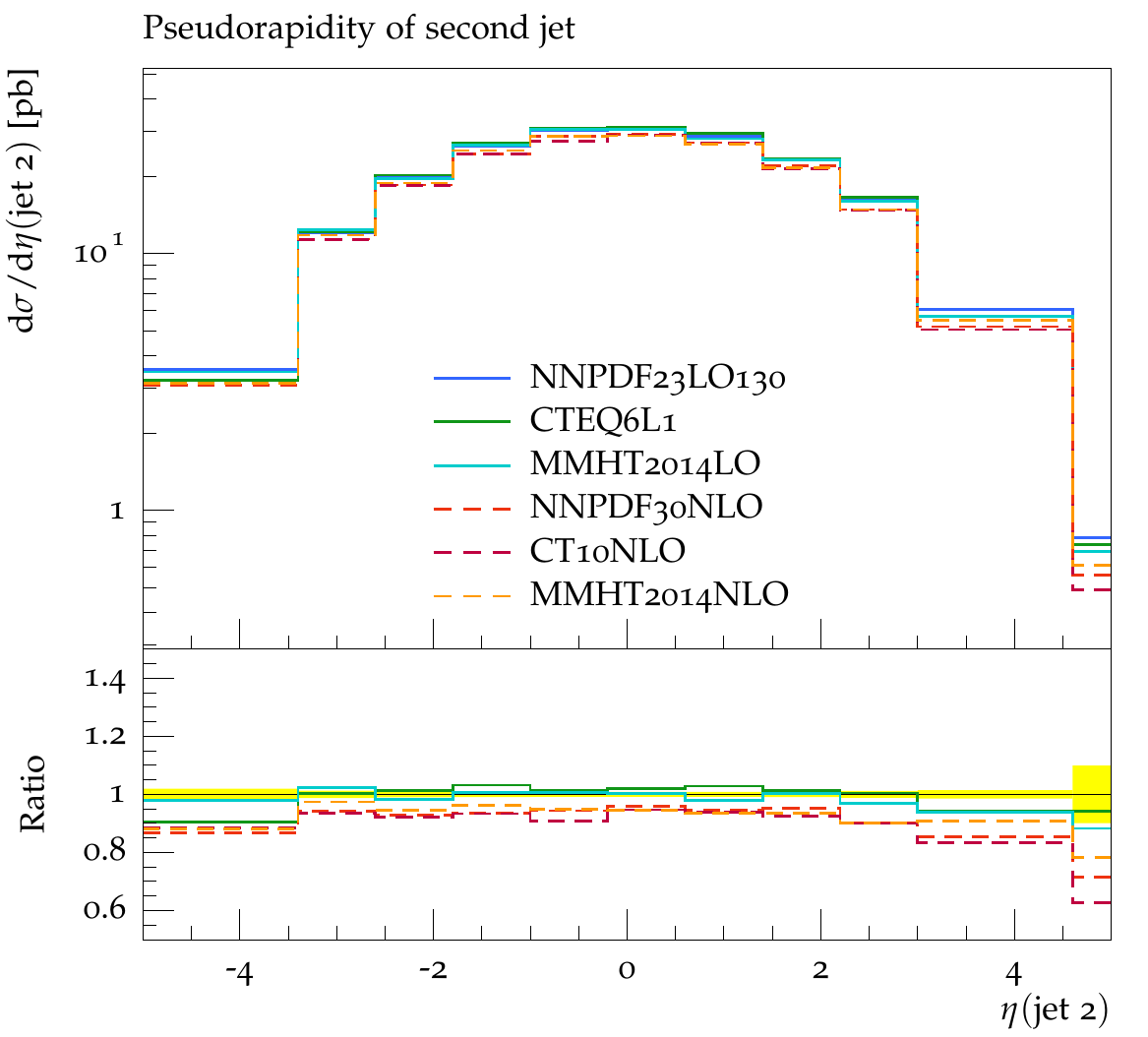}\\
  \img[0.48]{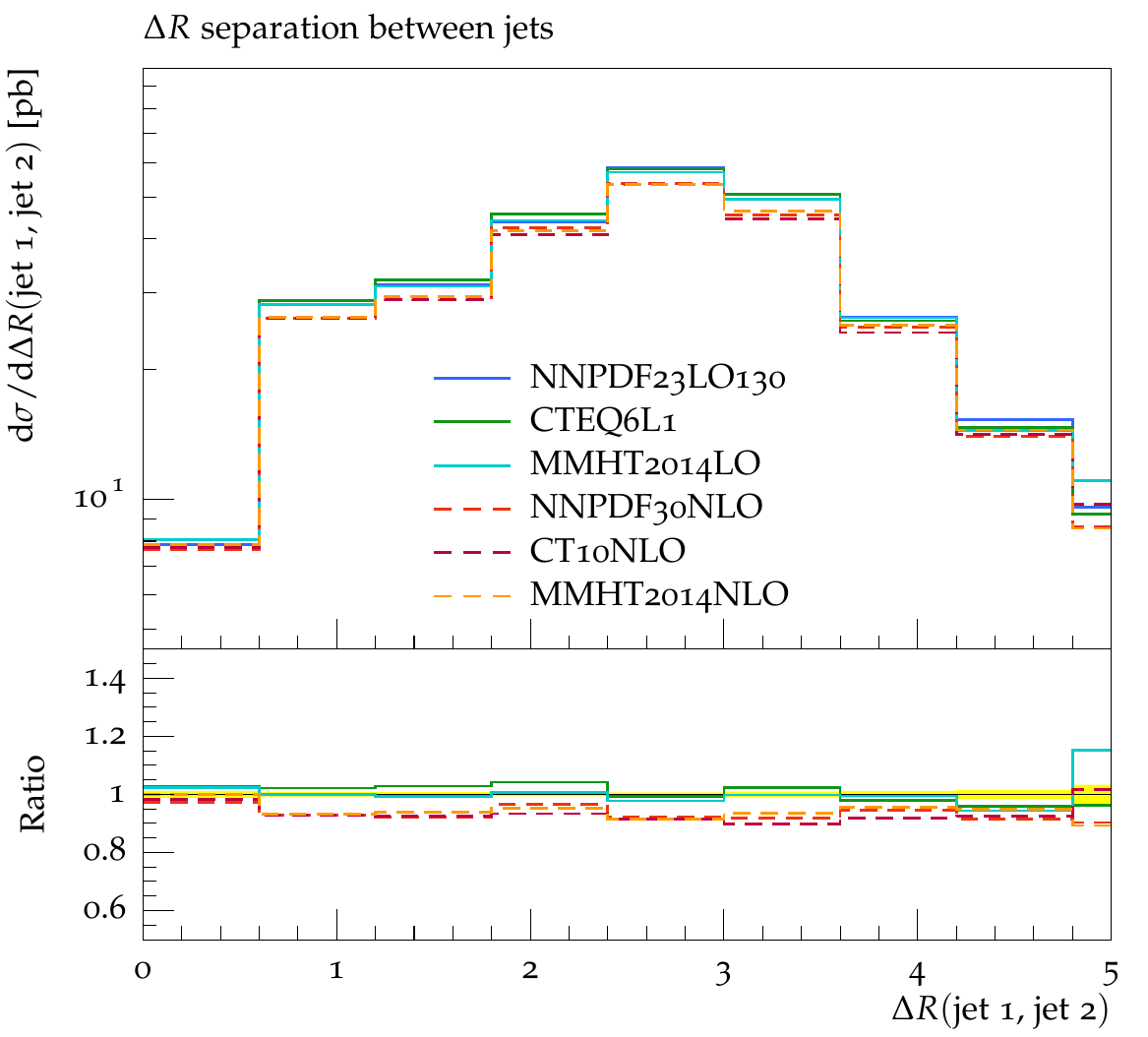}
  \img[0.48]{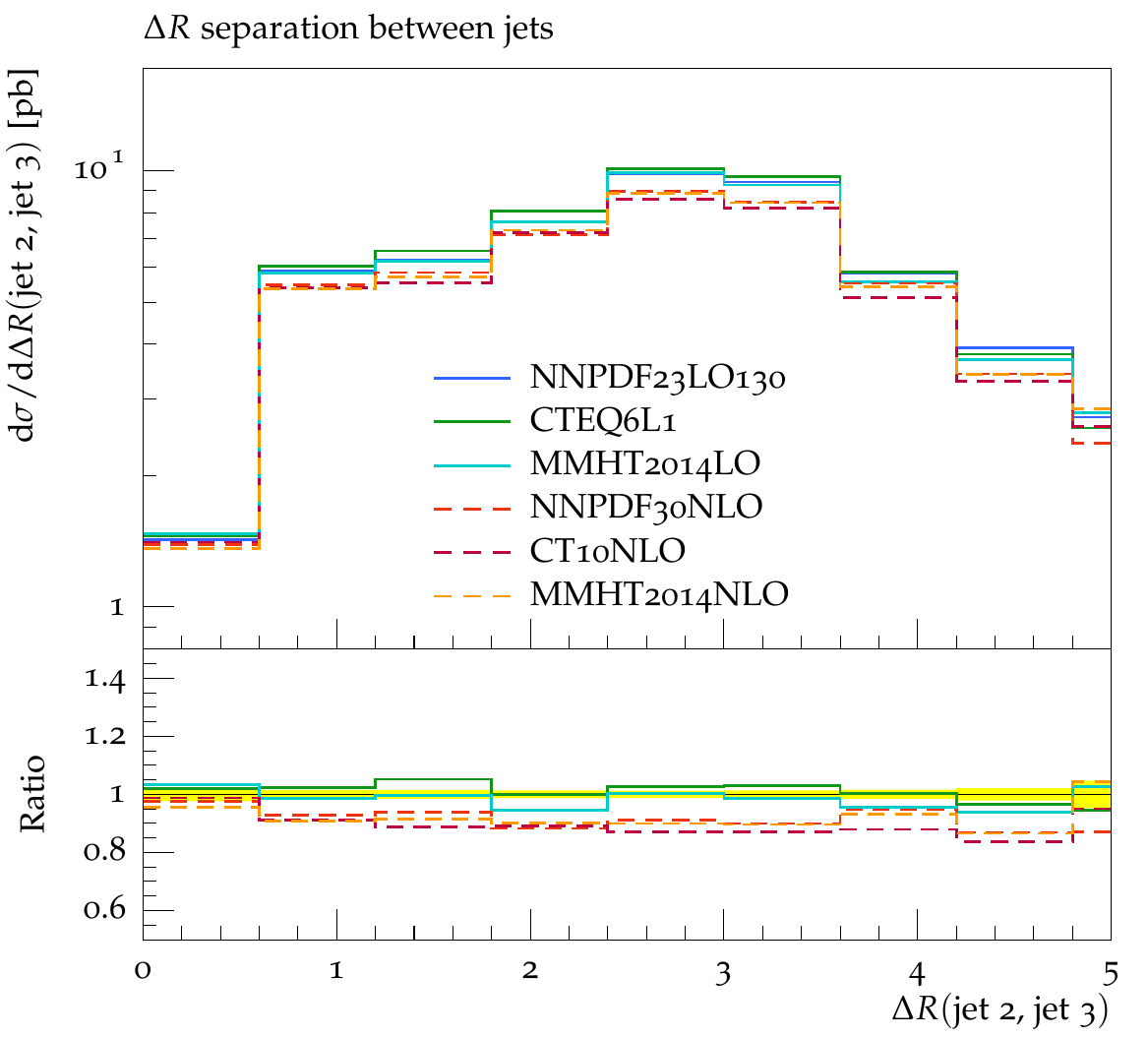}\\
  \img[0.48]{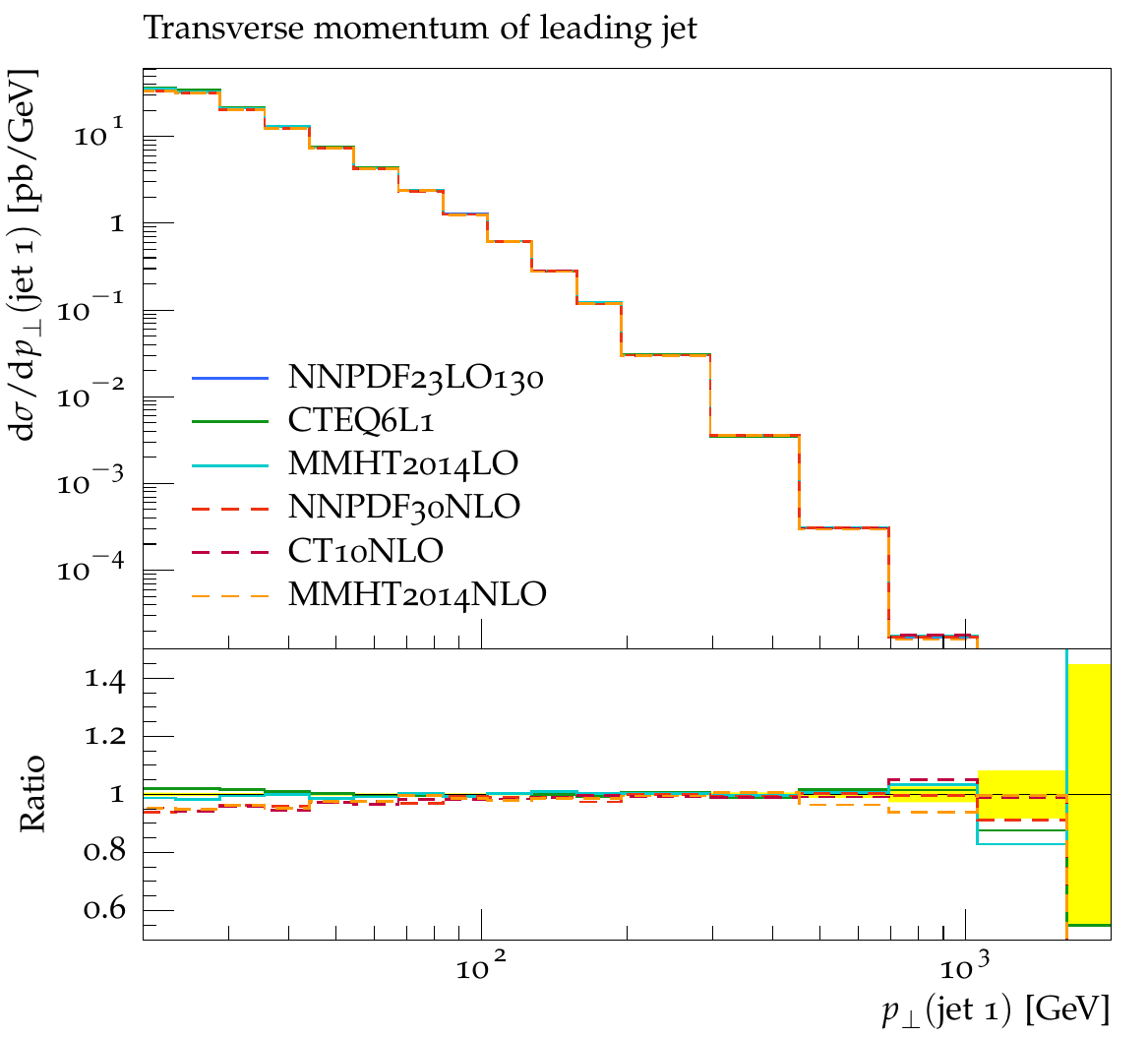}
  \img[0.48]{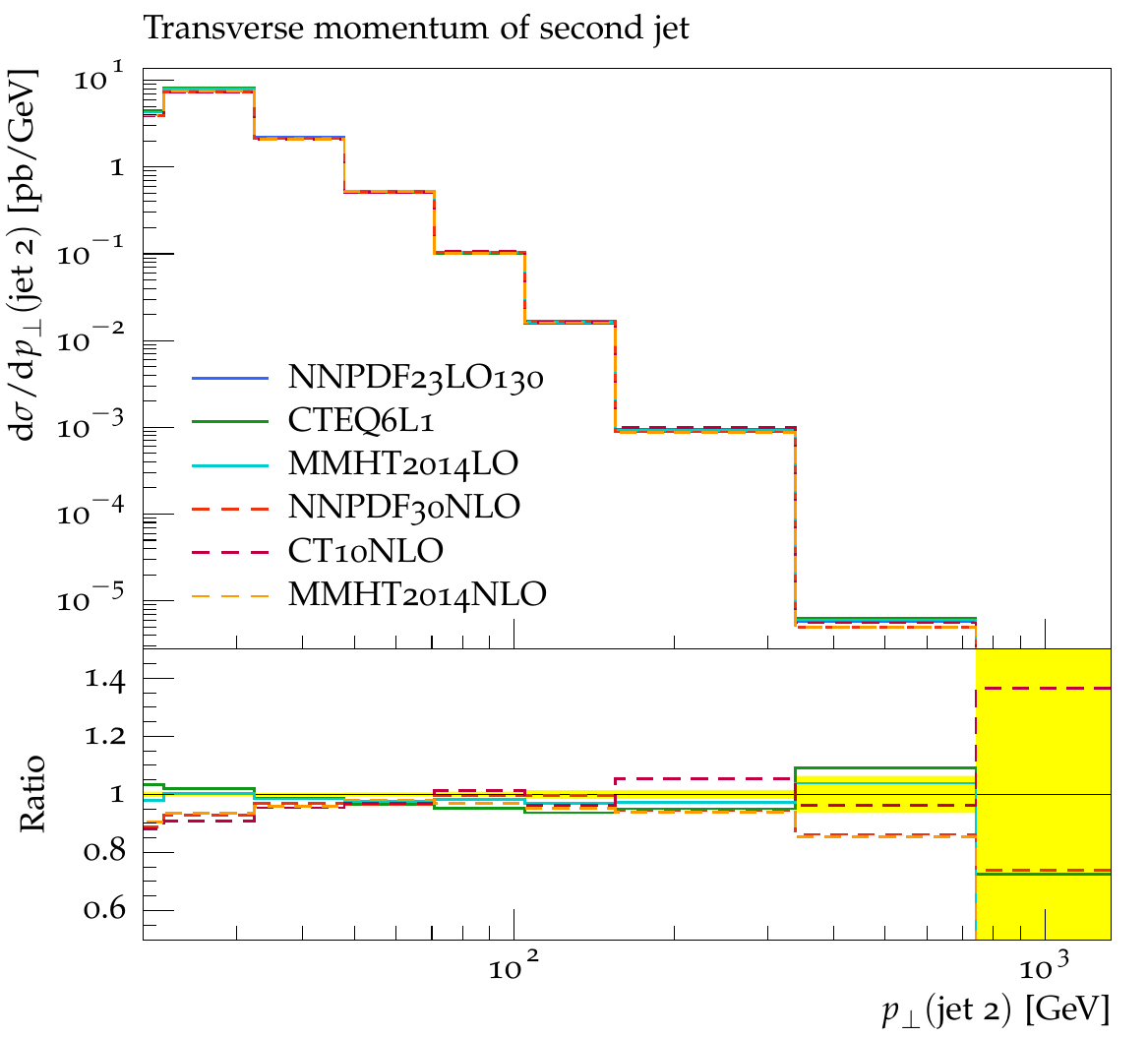}\\
  \caption{\Pythia8 $W + \text{jet}$ $\eta$, $\Delta{R}$ and $\pt$ distributions.}
  \label{fig:py8wjetetaetc}
\end{figure}

\begin{figure}[tp]
  \centering
  \img[0.48]{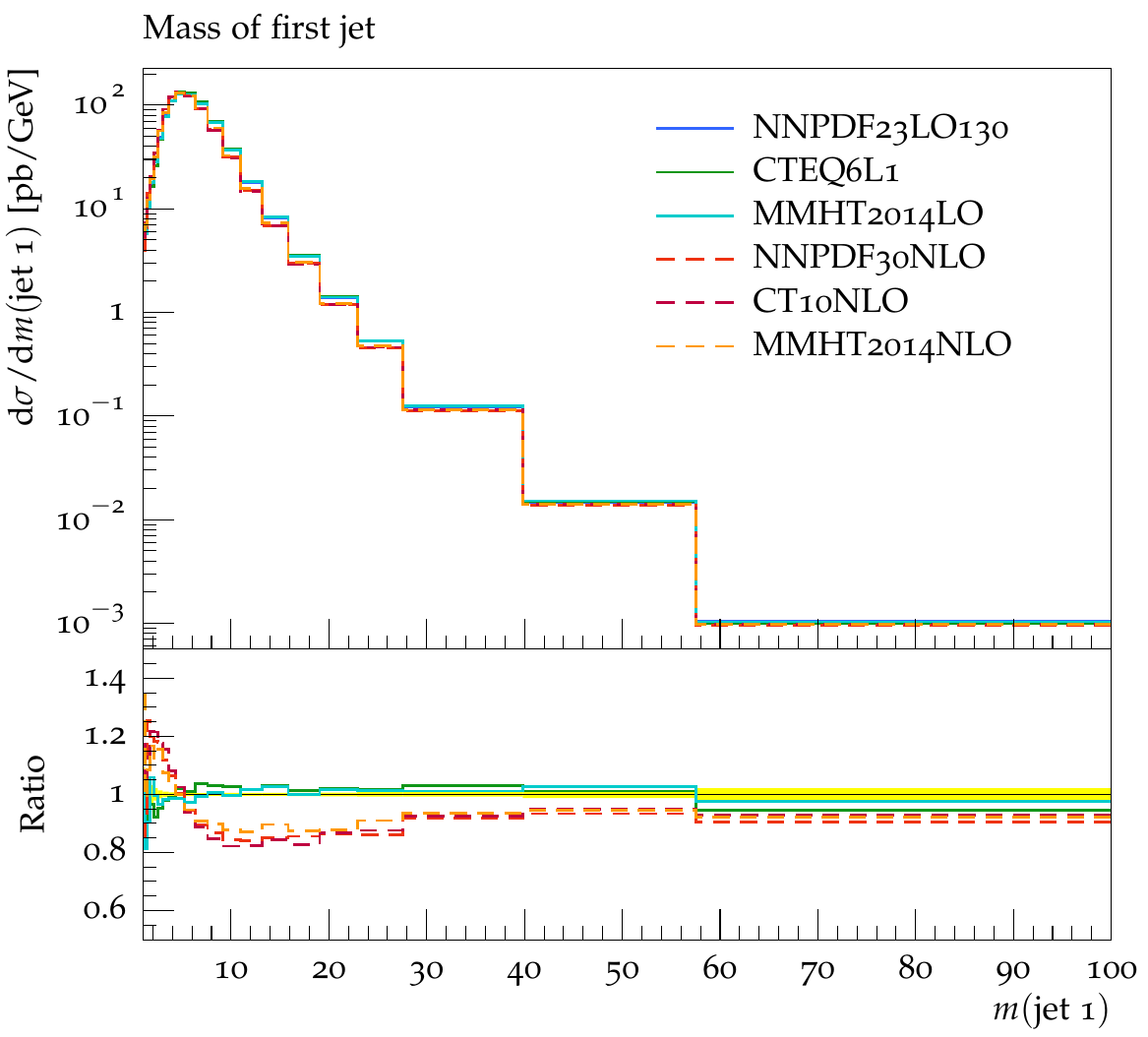}
  \img[0.48]{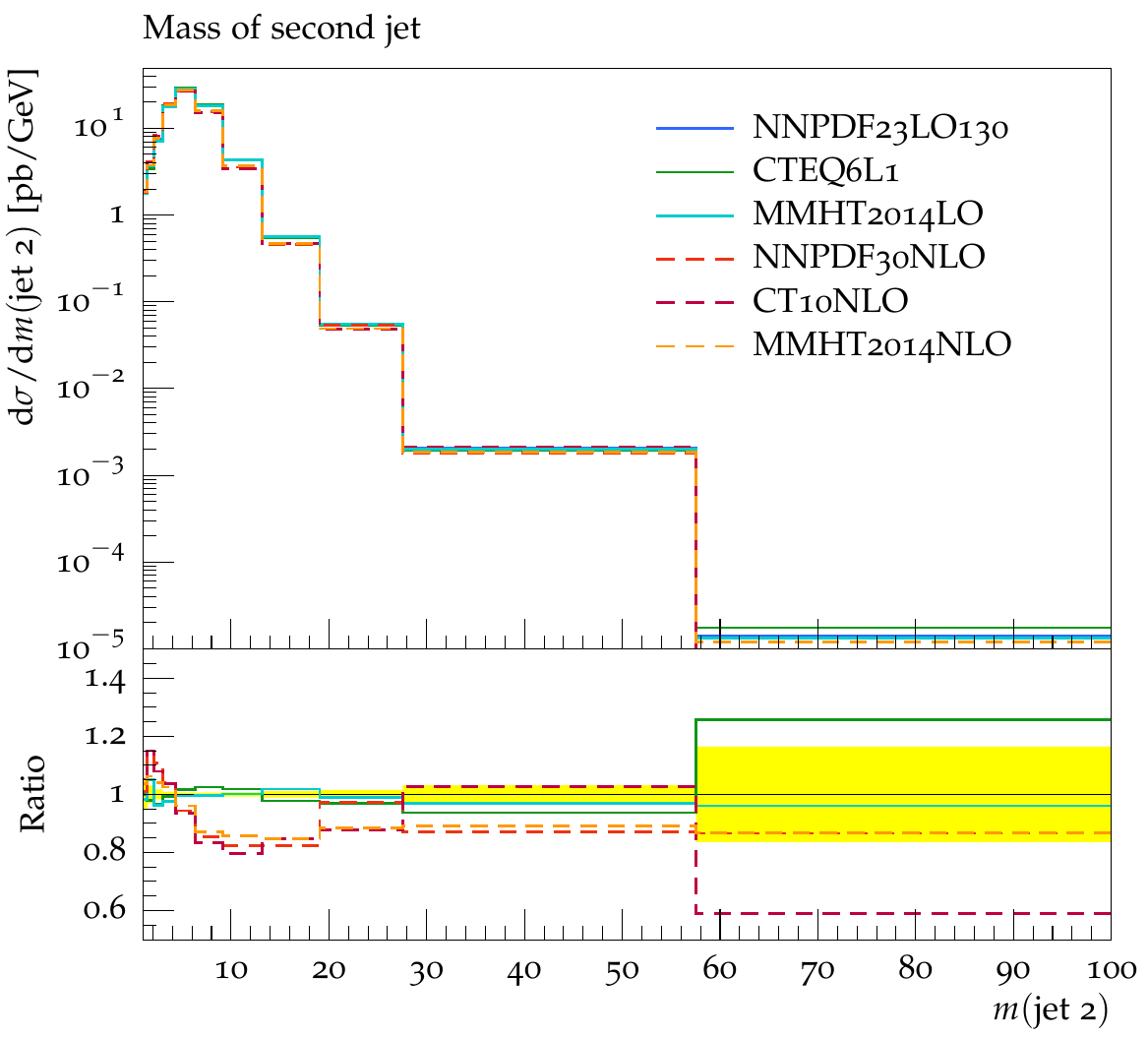}\\
  \img[0.48]{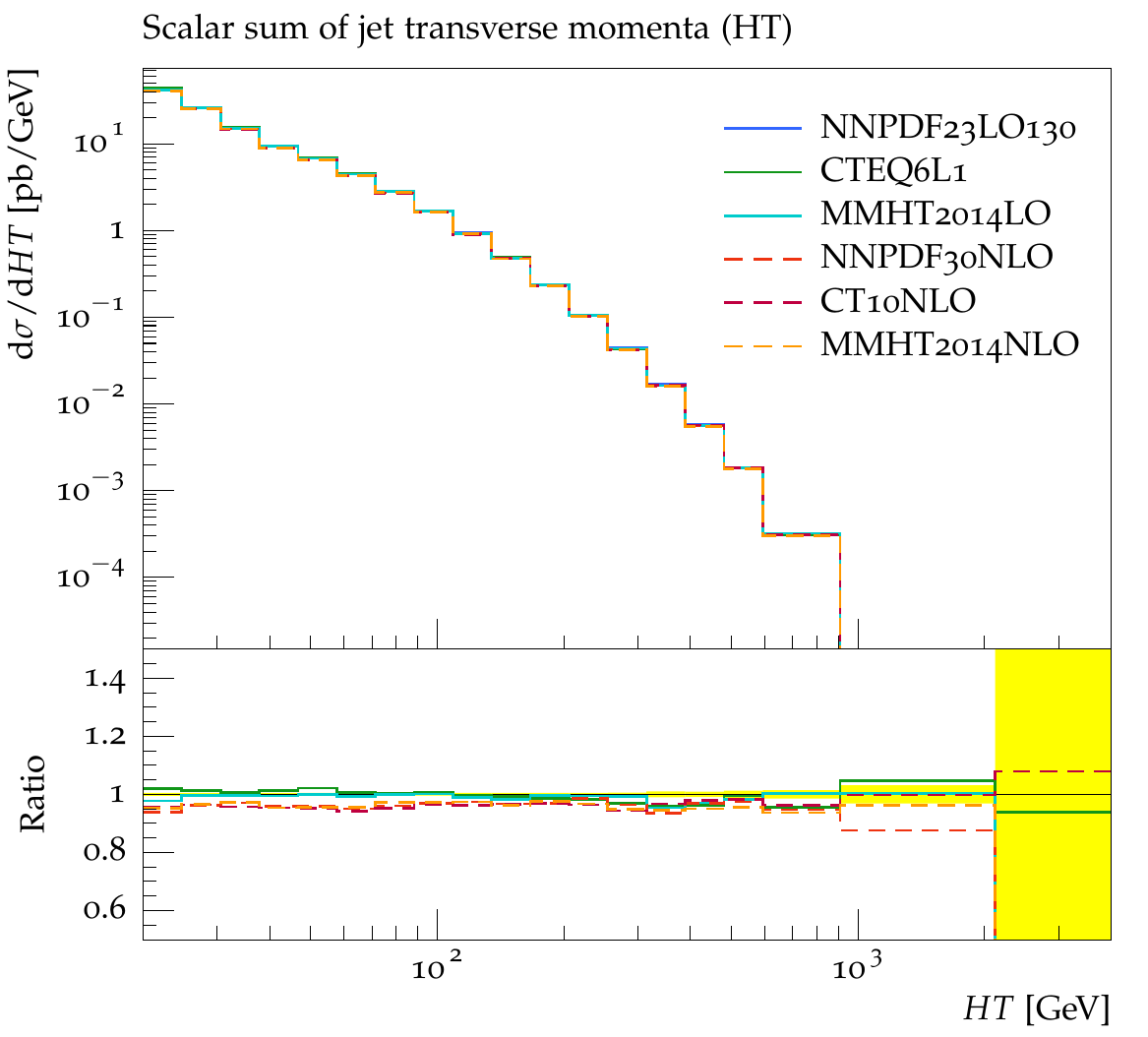}
  \img[0.48]{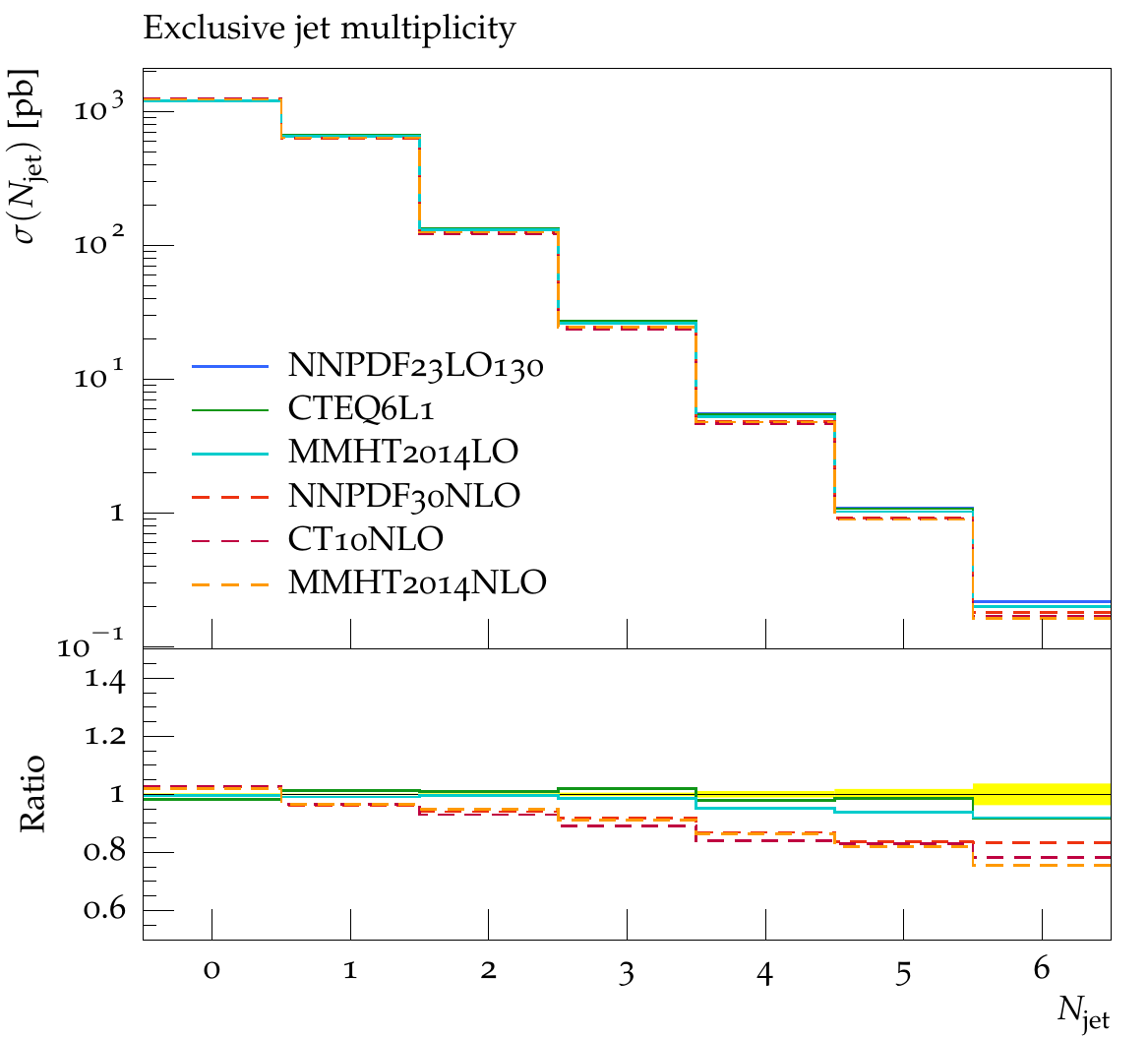}
  \caption{\Pythia8 $W + \text{jet}$ jet mass, $H_T$ and $N_\mathrm{jet}$ distributions.}
  \label{fig:py8wjetmassetc}
\end{figure}

2~million events were generated for each configuration, and were analysed
using the Rivet~\cite{Buckley:2010ar} \kbd{MC\_JETS} and \kbd{MC\_WJETS} validation
analyses, with jets defined by an anti-$k_T$ algorithm with $R=0.4$ and
$\pt > 20\,\GeV$. The results are shown in
Figures~\ref{fig:py8jets2eta}--\ref{fig:py8jets2mass} for the inclusive jets
process, and Figures~\ref{fig:py8wjetetaetc} and~\ref{fig:py8wjetmassetc} for
the inclusive $W+\text{jet}$ process. The LO and NLO soft modelling PDF choices are
respectively rendered as solid lines with blue/green colouring, and as dashed
lines with red/orange colouring to allow easy identification of the natural PDF
groupings.

These figures show several expected effects. First, the effects of switching
between leading order shower and MPI PDFs are small, on the roughly expected
scale of a few percent and less than 10\% in all statistically well-populated
regions of the plots. The use of NLO PDFs in shower \& MPI modelling produces
large differences with respect to the LO baseline setup, with variations of up
to 20\% in many observables, and rarely less than 10\%.

Secondly, the kinematics of objects such as the leading two jets in inclusive
jet production (which are dominated by the matrix element partons) are in fact
fairly stable even when NLO PDFs are used for showering, but extra jets in both
processes (dominated by initial state shower emissions) are more strongly
affected. The jet multiplicities are similarly strongly affected -- particularly
for the NLO shower PDFs at high multiplicities.

Further effects are seen when NLO PDFs are used for soft QCD simulation, notably
the strong reduction of multi-jet rates and the 20\% difference in jet masses
below $\sim 30\,\GeV$. These motivate further discussion of \alphaS treatment
and multi-parton interactions modelling, and are considered in the next section.

Broadly, these empirical data support several conventional rules of thumb:
\begin{itemize}
\item the effects of the particular leading-order parton shower PDF choice
  (without shower retuning) are limited to a few percent when considering
  perturbative QCD objects in the bulk of phase space;
\item reversing this observation, providing specific retunes of parton showers
  for different LO matrix element PDFs can only provide benefits of at most a few
  percent in some phase space regions;
\item we cannot directly conclude from this data about shower-PDF sensitivity in
  event simulation with NLO matrix elements, but as it is well-known that LO
  Pythia\,8 configurations can perform well in connection with NLO hard process
  events from \textsc{Powheg-Box}~\cite{Alioli:2010xd} or
  aMC@NLO~\cite{Alwall:2014hca}, the potential performance gain is again
  expected to be on the few-percent scale.
\end{itemize}

\subsection{\alphaS (in)consistency}

So far we have only considered the effect of changing parton density values,
without regard for the strong coupling \alphaS, whose boundary conditions and
evolution are linked to the PDF evolution.

\begin{figure}[tp]
  \centering
  \img[0.5]{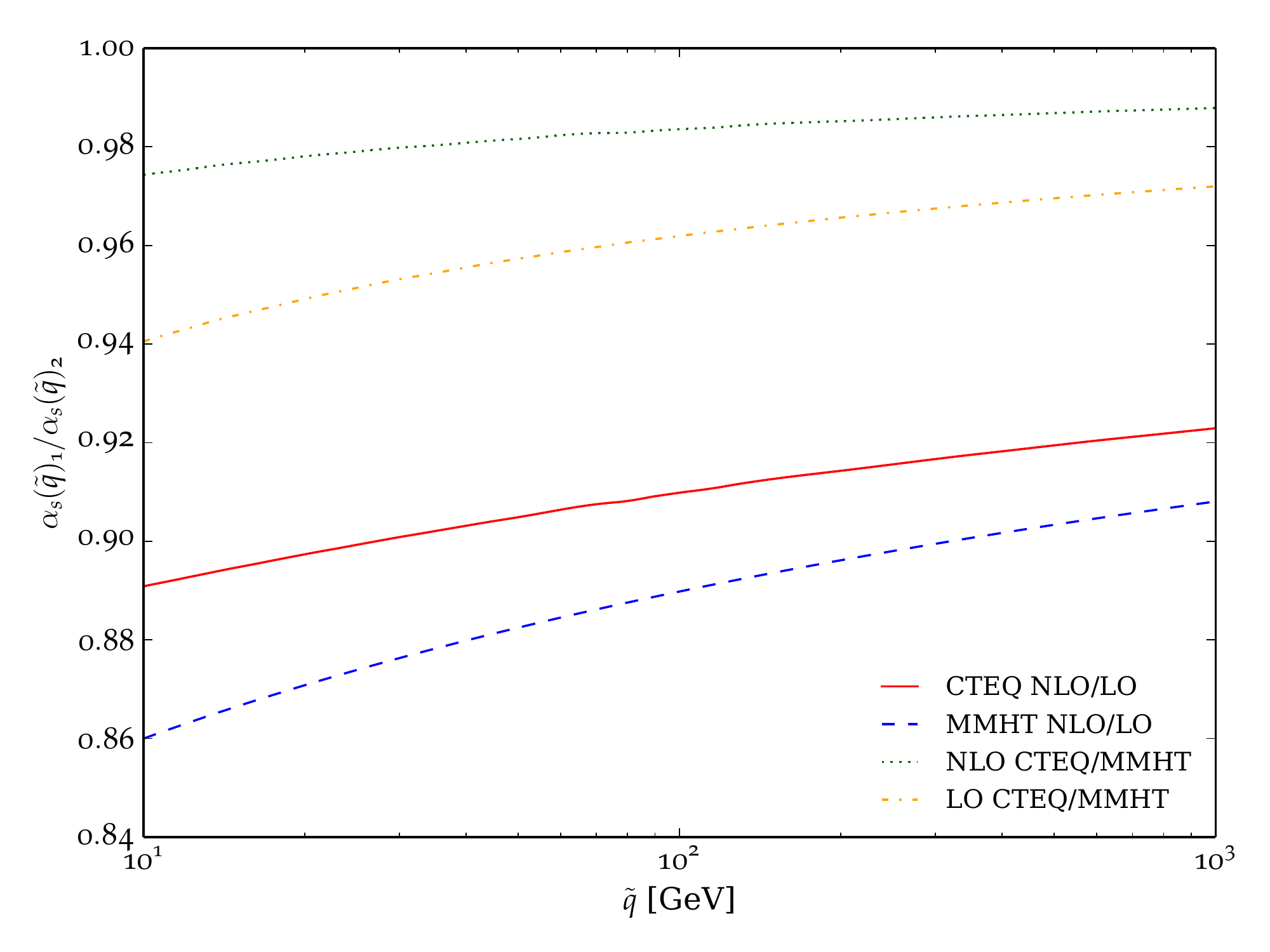}
  \caption{PDF \alphaS ratio distributions between CT10~NLO
    ($\alphaS(M_Z) = 0.118$) \& CTEQ6L1~LO ($\alphaS(M_Z) = 0.130$), and
    MMHT14~NLO ($\alphaS(M_Z) = 0.120$) \& LO ($\alphaS(M_Z) = 0.135$).}
  \label{fig:alphasratio}
\end{figure}

\subsubsection{PDF \alphaS ratios}

Ratios of \alphaS between PDF families and LO/NLO are shown in
Figure~\ref{fig:alphasratio} for a wide range of $\tilde{q}$ scales from
10--1000\,\GeV. Unsurprisingly, the biggest differences (up to 15\%) are between
LO and NLO PDFs' \alphaS values; and similarly unsurprisingly the second biggest
differences are between the two LO sets. The differences between NLO PDF \alphaS
values are a relatively flat $\sim 2\%$ across the scale range. For many
purposes this means that reweighting NLO MC samples between two NLO PDFs with
not-exactly-matching PDF couplings will induce acceptably small deviations from
a full resimulation\footnote{LHAPDF\,6 warns by default
  if reweighting is attempted between PDFs whose $\alphaS(Q)$ differ by
  more than 2\%.}. Between LO samples the \alphaS values may vary more
significantly, and whether ignoring this effect is acceptable depends on both
the PDFs and generators involved. LO \Sherpa, \Herwig++, \MadGraph and \Alpgen
all do use the PDF \alphaS in their matrix element evaluation, but Pythia\,8
does not (the fixed ME default is $\alphaS(M_Z) = 0.130$, the same as most LO
PDFs) and hence there may be no effect at all due to an \alphaS mismatch in
reweighting\footnote{Although use different tunes for the two PDFs may well
  change the ME \alphaS parameter -- \textit{caveat emptor}.}. Reweighting, and
indeed any attempt at naive exchange, between LO and NLO PDFs is very strongly
discouraged!

\subsubsection{\alphaS effects in lowest-order simulation}

Unlike its predecessor, version~8 of the Pythia parton shower generator does not
automatically change the \alphaS values in its parton showers (or the matrix
element, cf. above) to match that of the PDF used in initial-state showering;
instead the user is left free to choose whichever $\alphaS(M_Z)$ values (and
evolution orders) they like for not only the ISR and FSR showers independently,
but also the MPI matrix element and hadronization. The Herwig++ generator offers
similar freedom, and of the three most commonly used shower generators, only
Sherpa enforces full \alphaS consistency.

This freedom to use different \alphaS values in different parts of MC
simulations gives MC tuning studies extra degrees of freedom to best describe
observables (such as jet shapes) which should fall within the range of validity
of the parton shower approach. The validity of this flexibility is argued from
the limited (leading logarithmic) predictivity of current parton shower
algorithms, which is most important for lowest-order simulation such as the
Pythia\,8 built-in hard processes. The use of this approach in MC tuning has
motivated the approach taken here, where we have considered purely the effect of
the PDF itself and not the associated QCD parameters.

This is perhaps surprising, since na\"ively the parton shower \alphaS
configuration must match that of the hard process PDF and matrix element, to
avoid inconsistency. Unlike PDF values themselves, there is no ``double ratio
protection'' for \alphaS mismatches in parton shower evolution. But for
``Born-level'' leading-order simulation this inconsistency can again be absorbed
into the leading-log uncertainty of the shower formalism, allowing the mismatch
to be countered by other tunable shower parameters such as flexibility in
evolution start \& cutoff scales: historically this is exactly what has been
done to achieve acceptable data description.

In Figures~\ref{fig:py8jets2mass} and~\ref{fig:py8wjetmassetc} the 20\% lower
rate of multijet production with NLO shower PDFs seems consistent with the
significantly lower $\alphaS(M_Z)$ used in those PDFs, until one realises that
this is a purely PDF effect and that re-achieving good data description through
shower tuning with an NLO shower PDF would require use of extremely high shower
\alphaS values -- into a debatably unphysical regime with even larger ME/shower
\alphaS mismatches than already gave concern.

It is also worth noting that the large values of
$\alphaS(M_Z) \sim 0.13\text{--}0.14$ often resulting from parton shower tuning
are not badly inconsistent with the \alphaS values used in leading-order PDFs --
one of many arguments in favour of using LO PDFs for LO matrix elements, at
least at lowest order if not for multi-leg hard processes.


\subsubsection{Beyond Born-level}

The rise of more complete QCD calculations has cast a spanner into this
machinery, however. In NLO and multi-leg LO ME/PS matching, it is key that
parton emissions from shower or matrix element can be considered as
interchangable, with the additional (beyond Born) partons from the matrix
element being neatly assimilated into the shower evolution as improved splitting
functions. An inconsistency in ME and shower coupling breaks this equivalence to
some degree, leading to perverse effects such as the increased/reduced hardness
of \textsc{Alpgen} events when the shower \alphaS is respectively
reduced/increased~\cite{Cooper:2011gk}. While this is not directly a PDF effect
in the parton shower, the strict requirement of PDF/ME consistency in \alphaS
and the matching requirement of interchangable shower and ME partons.

In practice this effect is often tolerable for LO matched simulations --
especially when traded against the pragmatic gains from tuning the LL shower
model to data -- due to the similarity of \alphaS values in LO PDFs and in tuned
shower configurations. But it should be far less ignorable if either NLO PDFs
are used for the LO multi-leg ME calculation 
or for NLO matched configurations (obviously also using NLO PDFs), since the NLO
$\alphaS \sim 0.118$ value is much smaller than the ``natural'' value in the
parton shower. It is surprising, then, that in general matched NLO simulation
tunes have not paid great attention to this issue -- however, unambiguous NLO
matching configurations have yet to be established for
\textsc{Powheg-Box}~\cite{Alioli:2010xd} or aMC@NLO~\cite{Alwall:2014hca}, the
NLO matched generators (unlike Sherpa) for which the parton shower configuration
is not fully constrained, and hence for now we may be conflating \alphaS
consistency issues into that larger issue of configurational optimisation.

We conclude by noting two potentially useful tools in the MC simulation kit:
first, the argument from Catani, Marchesini \& Webber~(CMW)~\cite{Catani:1990rr} that for
consistency with full analytic resummation calculations, the \alphaS in parton
showers should be modified -- in the original formulation by
$N_\mathrm{f}$-dependent upward rescaling of $\Lambda_\mathrm{QCD}$. This
provides a motivation for using a shower \alphaS larger than that in the PDF \&
matrix element, but it remains unclear how this can be made consistent with the
requirements of matching schemes.  And secondly, an ``LO'' PDF with an
``NLO-like'' $\alphaS = 0.118$ already exists -- the CT09MCS
set~\cite{Lai:2009ne}. Whether this is a panacea for NLO matching configurations is
not clear, however, since previous use of ``modified LO'' PDFs -- albeit from
the MRST family rather than the CTEQ one -- led to unwanted artefacts in soft
event features such as underlying event: explicit study of this configuration is
needed to understand whether such effects are induced at a problematic level by
the CT09MCS set.

In summary, at lowest order there is significant freedom to choose special
\alphaS values for parton shower evolution without mismatches to the PDF
coupling inducing anomalous behaviours more significant than the intrinsic
uncertainty of the shower formalism. Once ME/PS matching is involved, things
become more complicated, but for multi-leg LO simulation the freedom is again
tolerable because of the substantial scale uncertainty of the ME and shower and
because typical shower values for \alphaS are not far from those used in LO PDF
fits. Matched NLO calculations, and particularly state-of-the-art 
multi-leg-NLO ones, are now the front-line in this long-running battle between
the ugly pragmatism of tuning and the theoretical requirements of full QCD
consistency -- the state-of-the-art hence may not describe data as well as LO+LL
simulations tuned to data, but the increasing ability to calculate QCD processes
from first principles (and accordingly predictivity in so-far unmeasured phase
spaces) can only be welcomed.


\subsection{PDFs in multi-parton interactions}

The demonstrations and arguments above make a compelling case that we should not
be overconcerned about PDF consistency between PDFs \& matrix elements and
parton showers, because of the relative stability of the PDF ratios which appear
in the ISR Sudakov form factor, eq.~\eqref{eq:sudakov}.

But a surprising effect in this Pythia\,8 study is seen in the jet mass plots of
Figures~\ref{fig:py8wjetmassetc} and \ref{fig:py8jets2mass}. Given that the the
shower couplings are not automatically affected by the shower PDF, and that
perturbative jet mass dominantly arises from the broadening effects of the
\emph{final-state} shower (whose Sudakov factor includes no PDF terms at all),
the observed large effect seems perverse. But we have also to consider the
effects of PDFs in multi-parton interactions (MPI).

It is well known that as the minimum parton/jet scale, $\hat{p}_T^\mathrm{min}$,
is reduced in jet cross-section calculations, the resulting cross-section
diverges and eventually exceeds the total $pp$ cross-section. In the usual
eikonal approach to MPI modelling, the ratio of calculated partonic to hadronic
collision cross-sections -- i.e. the factor by which unitarity is naively
violated -- is interpreted as the mean number of partonic interactions in each
$pp$ collision, $\mu_\mathrm{int}$. Each $pp$ event then samples from a Poisson
distribution with mean $\mu_\mathrm{int}-1$ to choose how many soft QCD partonic
interactions, $n_\mathrm{int}$, will accompany the hard scatter. Since
$\mu_\mathrm{int}$ depends on the partonic cross-section at low-$p_T$, which is
sensitive to low-$x$ PDFs, changes of PDF can have a very significant effect on
levels of MPI activity (i.e. underlying event), even when restricted to LO fits
as is recommended~\cite{Sjostrand:2014mcpdf}.

It is hence most plausible that underestimation of the level of partonic
multiple scattering, in particular due to the smaller low-$x$ gluon in NLO PDFs,
and the absence of a retune of MPI model cutoffs and suppression factors is
responsible for the large effects of ``shower'' PDF choice on jet mass
observables in Pythia simulation.

A switch of PDF in the ``soft'' (i.e. non-hard-process) components of an event
generator will hence in general need to retune the MPI model. This may be
necessary even if restricting to LO PDFs with the same \alphaS, since different
PDF families can have significantly different low-$x$ gluon distributions for
the same strong coupling: the changes in Figures~\ref{fig:py8jets2mass}
and~\ref{fig:py8wjetmassetc} are not always negligible within the LO group,
especially below $m \sim 30\,\GeV$. Luckily, previous tuning experience with the
Pythia model has shown that the maximum ``plateau'' level of MPI activity can be
adjusted with minimal impact on more detailed underlying event observables, or
the parton shower, via a simple 1-parameter tune of the $p_T^0$ screening
scale. It is hence possible to switch soft-process PDFs with only a simple --
perhaps by-hand -- retuning, rather than needing to re-employ the more
comprehensive machinery needed for a full shower+MPI generator tune.

\section{Summary}

This note has presented empirical evidence for the usual approaches to
reweighting and retuning Monte Carlo parton shower event generators for use with
different PDFs, along with discussion of some of the observable artefacts and
physical principles involved. We hope these plots and discussion will prove
useful, particularly as a pedagogical introduction and to highlight areas where
a 100\% satisfying solution to shower configuration ambiguities has yet to be
found.

Several popular rules of thumb are supported by this
presentation. 
For unambiguously leading-order hard process simulation, the use of LO PDFs is
recommended not only because of the compensation for missing LO matrix element
effects built into the PDF fit but also for closer equivalence of the PDF
\alphaS value to the values typically obtained in parton shower tuning to
$e^+e^-$ and hadron collider data. PDF reweighting between LO and NLO sets --
thankfully rare in practice, although not unheard of -- is seen to induce severe
effects due to both PDF and \alphaS effects and is strongly discouraged.

The smallness of differences between LO and particularly between NLO PDF ratios
and the corresponding \alphaS suggest that reweighting at ME-level only, and
hence neglecting their effects on parton showers, should be an acceptable
approximation -- provided that the neglected effects will be covered by the
systematic uncertainties inherent in the parton shower formalism. Explicit
simulation confirms this, with neglected deviations within the group of major LO
PDFs being restricted to $\sim 5\%$ at most. Even observables sensitive to
multiple initial state emissions, such as $H_T$ do not show more significant
effects. This argument in favour of ``na\"ive'' reweighting can also be used to
justify switching of shower PDFs in generator configurations without a need for
detailed shower parameter retuning -- although the typically associated change
of multi-parton interactions PDF will necessitate some simple tune modification
to account for sensitivity to low-$x$ parton distribution differences. The
relative insensitivity of parton showers to PDFs is good news insofar as it
suggests that a single shower MC tune made with a given leading-order
PDF/ME/shower/MPI PDF can be reused with many different hard-process PDFs rather
than needing a family of tunes to cover all hard process simulation options.

More substantial than PDF values themselves may be the corresponding variations
in the strong coupling, \alphaS. If propagated to the parton shower, \alphaS
changes can have a significant effect on many observables, particularly at
LO. While generator codes may allow for inconsistency in both PDF and \alphaS
treatment, this pragmatic freedom can easily prove counterproductive once
higher-order hard process calculations are used -- as is these days the norm for
simulation of Standard~Model processes.  The CMW~\cite{Catani:1990rr} argument from
resummed calculations for \alphaS enhancement in parton showers has some
purchase but does not obviously play well with the need for equivalence of
shower \& ME parton emission in ME/PS-matched calculations. The convenience of
PDF reweighting and weight-based systematics without explicitly incorporating
the effects on parton showers is for now a clear win for pragmatism, but as hard
process modelling becomes more and more sophisticated we must ensure to monitor
its evolving validity.

\vspace*{1em}
\section*{Apologies}
\emph{My apologies to all those to whom I promised I would write up these
  studies more than 18 months ago. I hope late really has proven to be better
  than never!}
\vspace*{1em}

\bibliographystyle{atlasnote}
\bibliography{refs}





\end{document}